%% file: main.tex
\documentclass[sigconf]{acmart}
\AtBeginDocument{%
  }

\copyrightyear{2026}
\acmYear{2026}
\setcopyright{cc}
\setcctype{by-nc-nd}
\acmConference[CHI '26]{Proceedings of the 2026 CHI Conference on Human Factors in Computing Systems}{April 13--17, 2026}{Barcelona, Spain}
\acmBooktitle{Proceedings of the 2026 CHI Conference on Human Factors in Computing Systems (CHI '26), April 13--17, 2026, Barcelona, Spain}
\acmDOI{10.1145/3772318.3790785}
\acmISBN{979-8-4007-2278-3/2026/04}

\usepackage{algorithm} 
\usepackage{subcaption}
\usepackage{amsmath}

\usepackage{multirow}

\usepackage[utf8]{inputenc}
\usepackage[T1]{fontenc}        

 \newcommand{\my}[1]{}
 \newcommand{\ignore}[1]{}

\usepackage{enumitem}
\usepackage{soul} 
\usepackage{color} 
\soulregister\cite7
\soulregister\ref7 
\usepackage{hyperref}   
\usepackage[capitalise]{cleveref} 
\newif\ifhighlighton

\highlightontrue   
\highlightonfalse
\ifhighlighton
\else
  
\fi

\begin{document}

\title{Understanding the Effects of AI-Assisted Critical Thinking on Human-AI Decision Making}

\author{Harry Yizhou Tian}
\email{tian253@purdue.edu}
\affiliation{%
  \institution{Purdue University}
  \city{West Lafayette}
  \state{Indiana}
  \country{USA}
}

\author{Hasan Amin}
\email{hasanamin@purdue.edu}
\affiliation{%
  \institution{Purdue University}
  \city{West Lafayette}
  \state{Indiana}
  \country{USA}
}

\author{Ming Yin}
\email{mingyin@purdue.edu}
\affiliation{%
  \institution{Purdue University}
  \city{West Lafayette}
  \state{Indiana}
  \country{USA}
}

\renewcommand{\shortauthors}{Tian et al.}
\newcommand{\aact}{AACT}
\newcommand{\AACT}{AIAssistedCriticalThinking}



\input{sections/abstract}

\maketitle

\input{sections/1_intro}

\input{sections/2_background}

\input{sections/3_system}

\input{sections/4_experiment}

\input{sections/5_results}

\input{sections/6_discussion}

\input{sections/7_conclusion}

\begin{acks}
We thank the support of the National Science Foundation under grant IIS-2229876 and IIS-2340209 on this work. 
We are very grateful for the reviewers' feedback, which helped improve this paper significantly.
Any opinions, findings, conclusions, or recommendations expressed here are those of the authors alone.
\end{acks}

\bibliographystyle{ACM-Reference-Format}
\bibliography{refs}

\newpage
\appendix

\input{sections/appendix}

\renewcommand{\thefigure}{\thesection.\arabic{figure}}
\renewcommand{\thetable}{\thesection.\arabic{table}}

\end{document}

%% file: sections/abstract.tex
\begin{abstract}

Despite the growing prevalence of human-AI decision making, the human-AI team’s decision performance often remains suboptimal, partially due to insufficient examination of humans’ own reasoning. In this paper, we explore designing AI systems that directly analyze humans' decision rationales and encourage critical reflection of their own decisions. We introduce the \textit{AI-Assisted Critical Thinking (AACT)} framework, which leverages a domain-specific AI model’s counterfactual analysis of human decision to help decision-makers identify potential flaws in their decision argument and support the correction of them. 
Through a case study on house price prediction, we find that AACT outperforms traditional AI-based decision-support in reducing over-reliance on AI, though also triggering higher cognitive load. Subgroup analysis reveals AACT can be particularly beneficial for some decision-makers such as those very familiar with AI technologies.  
We conclude by discussing the practical implications of our findings, use cases and design choices of \aact{}, and considerations for using AI to facilitate critical thinking.

\end{abstract}

%% file: sections/1_intro.tex
\section{Introduction}
\label{sec:intro}

Artificial intelligence (AI) is increasingly deployed to assist human decision making in high stake domains such as criminal justice \citep{Dodge2019, yacoby2022}, academic admissions \citep{zhang2023admission}, recruiting \citep{amazon}, and medical diagnosis \citep{cai2019medical, lee2021medical, chen2025medical}. In this \textit{human-AI decision making} paradigm, AI acts as decision-support while human decision-makers make their final decision. However, research found that humans tend to engage superficially towards the AI information \cite{cff, Rastogi2022anchor, bansal2021}, resulting in a lack of deeper thinking, and consequently inappropriate reliance on AI \citep{cff, vasconcelos2023explanations, de2025cognitive, lai2019, wang2021} and suboptimal human-AI decision-making performance \citep{bansal2021, schemmer2022should, vaccaro2024combinations, steyvers2024three}. Recognizing these limits, researchers have begun to design AI-based decision systems that encourage deeper engagement. For example, \textit{cognitive forcing functions} compel independent reasoning before advice is revealed \citep{cff, de2025cognitive}. ``\textit{Hypothesis-driven XAI}'' presents evidence supporting and challenging different decisions instead of using AI decision as recommendation \citep{Miller2023}. Other studies have examined re-framing AI information from recommendation statements to reflective questions \citep{danry2023}, devils' advocate \citep{chiang2024, ma2024-3roles}, or constructive dialogue between human and AI \citep{ma2025, reicherts2022extending, reicherts2025ai}.

While these approaches encourage more deliberation and reflection, they share a common limitation: they focus on promoting reflection around \textit{the AI's outputs}, but offer little direct, actionable support for improving the human's critical reflection of \textbf{\textit{their own reasoning}}. 
This is because most interventions remain organized around the \textit{AI's perspective}: cognitive forcing functions shape when or how much AI advice is revealed  \citep{cff}; 
hypothesis-driven XAI \cite{Miller2023} and devil's advocate designs \cite{chiang2024} vary the evidence the AI provides;  
and question-based \cite{danry2023} or dialogue-based interfaces \cite{ma2025, reicherts2022extending, reicherts2025ai} guide users through queries that trace the AI's own reasoning trajectory.  
Across these approaches, the internal reasoning processes that humans rely on when forming a decision (i.e., their decision rationale) are rarely elicited or modeled. 
Even when humans' reasoning processes are solicited (e.g., in the human-AI deliberation framework~\cite{ma2025} or the Socratic deliberation method~\cite{khadar2025wisdom}), they are used to surface disagreements between humans and AI or to prompt large language model interrogations that rely on broad world knowledge. What remains missing is a domain-grounded, structured analysis of the human's reasoning that can inform targeted, actionable critique and revision. 
As humans' internal reasoning often contains the very hidden assumptions, omissions, and inconsistencies that undermine decision quality, the absence of AI systems capable of providing rigorous, quantitative assessment of that reasoning rooted in the specifics of the decision context represents a major missed opportunity for improving human-AI decision making.

To address this gap, we propose the \textit{AI-Assisted Critical Thinking} (\aact{}) framework, which leverages AI to computationally analyze and improve \textit{human's own  decision making} by encouraging critical reflection on their reasoning processes.
We build \aact{} upon the Recognition/Metacognition model~\cite{cohen1996,cohen1998critical}, a framework for real-world time-stressed decision making, and focus specifically on assisting human decision makers in \textit{critiquing} and \textit{correcting} their own decisions and arguments.
In \aact{}, a domain-specific AI model that is trained for the target decision context engages in \textit{counterfactual perspective-taking} by adopting the human's perspective and performing counterfactual analyses over the human's decision arguments to identify meaningful critiques. This analysis enables the AI to surface potential flaws in the human's decision arguments, including incompleteness, unreliability, and conflict, and to prompt \textit{targeted self-reflections} of these flaws through counterfactual questioning.  
Moreover, \aact{} also supports humans in correcting these potential flaws 
through two mechanisms: \textit{AI-based correction suggestions}, which presents AI's suggestions in revising their decision arguments, and \textit{data-based triangulation}, which introduces insights from the AI's training data as external checks. 

To comprehensively evaluate the effects of AI-Assisted Critical Thinking on human-AI decision making, we instantiated the \aact{} framework as a conversational AI system and conducted a controlled user study on Prolific, 
using a house price prediction task as a case study. Within this study, we compared \aact{} against established human-AI decision making frameworks such as Explainable AI (XAI) and Hypothesis-driven XAI \cite{Miller2023},
examining outcomes such as decision accuracy, reliance on AI, task learning, and participants’ subjective perceptions of the AI assistance. 
Because people’s preferences for critical thinking may vary by context and user characteristics \cite{lai2011}, we further analyzed treatment effects across participant subgroups defined by factors such as domain knowledge and familiarity with AI.

Our results show that compared to traditional human-AI decision making frameworks, \aact{} can be particularly effective in \textit{reducing decision-makers' over-reliance on AI}, though it may also come at a cost of increased under-reliance on AI and heightened mental demand. 
Moreover, we also find \aact{} to be \textit{more beneficial for certain subgroups of decision-makers}---for example, \aact{} mostly helps reduce over-reliance in decision-makers who are more familiar with the decision domain, very familiar with AI, and more educated. 
Our exploratory qualitative analysis further suggests that \aact{} is often preferred by decision-makers that prioritize decision autonomy and reflective thinking.

In summary, we make the following contributions in this paper:
\begin{enumerate}[topsep=0pt, partopsep=0pt, itemsep=0pt, parsep=0pt]
    \item Theoretically, we adapt the Recognition/Metacognition (R/M) model to human-AI decision making to propose AI-Assisted Critical Thinking (\aact{}). Unlike prior reflection-oriented approaches that focus on AI outputs, 
    \aact{} provides domain-grounded, structured analysis of the human's own reasoning and supports targeted refinement of that reasoning.
    \item Technically, we instantiate \aact{} as a conversational AI system, 
    implementing its counterfactual reasoning components and structured critique-and-correction workflows through conversational user interfaces.
    \item Empirically, we conduct a controlled user study examining \aact{}'s effects on  human-AI team decision accuracy, appropriate reliance on AI, and cognitive load, and demonstrate heterogeneous impacts across participant subgroups.
    \item Practically, we discuss real-world use cases and design implications for leveraging AI to scaffold metacognitive reflection on users' own reasoning, complementing existing reflection-based decision-support paradigms.
\end{enumerate}

%% file: sections/2_background.tex
\section{Related Work}
\label{sec:background}

\subsection{Human–AI Team Decision Making}
AI systems are increasingly integrated into high-stakes domains such as medical diagnosis \citep{cai2019medical, lee2021medical, chen2025medical}, criminal justice \citep{Dodge2019, yacoby2022}, and academic admissions \citep{zhang2023admission}, creating a new paradigm of human–AI collaborative decision making. The promise of these systems is to combine human expertise with AI's analytical capabilities to achieve ``complementarity,'' wherein the human-AI team outperforms either human or AI alone in decision making. Yet, evidence shows that such synergy often fails to materialize \citep{bansal2021, schemmer2022should, vaccaro2024combinations, steyvers2024three}.  
A central obstacle is inappropriate reliance on AI advice. Overreliance occurs when humans follow incorrect AI predictions \citep{cff, vasconcelos2023explanations, de2025cognitive}, while underreliance describes ignoring correct AI suggestions \citep{lai2019, wang2021}. 
These problems are exacerbated by cognitive biases, such as anchoring \citep{tversky1974judgment, Rastogi2022anchor} and illusions of explanatory depth \citep{illusion}, that distort how users interpret AI outputs.  

One prominent approach to address reliance issues is to improve humans' understandings of AI advice through AI explanations, or \textit{Explainable AI}. Techniques such as LIME \citep{lime}, Anchors \citep{Rastogi2022anchor}, and Interpretable Decision Sets \citep{IDS} exemplify efforts to make model reasoning more transparent. 
However, systematic reviews and empirical studies highlight mixed effectiveness: explanations can help users predict model behavior \citep{hase2020} or improve calibration \citep{zhang2020}, but they can also mislead, act as ``placebos,'' or inflate confidence without increasing accuracy \citep{placebo, bucina2020, lai2021}. 
Their utility often depends on user expertise \citep{Szymanski2021expertise, ehsan2024xai}, task complexity \citep{salimzadeh2023missing}, and even framing effects \citep{danry2023}. 
Recent syntheses argue that explanations rarely yield true complementarity and call for task-sensitive evaluation of explanation utility \citep{fok2024search, chaleshtori2024evaluating}.

\subsection{Facilitating Critical Reflection in Human–AI Decision Making}
More recently, researchers have pointed out that a major limitation in existing human-AI decision making is that people tend to engage only superficially with AI outputs~\cite{cff, Rastogi2022anchor, bansal2021}, resulting in a lack of deeper thinking and inappropriate reliance towards AI~\cite{cff, vasconcelos2023explanations, de2025cognitive, lai2019, wang2021}. To address this, various approaches have been developed to foster more deliberate, reflective engagement and higher-order thinking~\cite{yatani2024, collins2024,lu2024does}. Broadly, these approaches fall into strategies that alter \textit{when or what AI outputs are shown}, or \textit{how AI information is communicated}, while in most cases reflection remains anchored on the AI's perspective rather than the \textit{human's own reasoning}.

\input{tables/background}

One category of reflection-provoking approaches varies \textit{when or what} AI advice is revealed. For example, ``cognitive forcing functions'' prompt human decision-makers to think independently~\cite{cff, de2025cognitive} by requiring unaided judgments before revealing AI advice or by offering only partial explanations~\cite{vasconcelos2023explanations, reicherts2022extending, drosos2025makes}. While effective in interrupting automatic acceptance of AI suggestions, they do not directly engage with the human's internal reasoning; they simply create conditions intended to trigger that reasoning. Another approach of ``hypothesis-driven XAI'' \cite{Miller2023} proposes using AI to surface evidence for and against multiple possible hypotheses instead of a single prediction. Studies show its effectiveness in reducing humans' over-reliance on AI \cite{le2024} and improving prediction accuracy when AI performance is low \cite{ma2024-3roles}, yet the reflection it encourages remains focused on  competing AI-constructed evidence rather than on examining the human's own decision argument. Likewise, the ``devil's advocate'' approach uses AI to provide opposing arguments that challenge the human's decision \cite{chiang2024,ma2024-3roles, lee2025conversational}, but these arguments are typically derived from the AI's independent analysis of the task, rather than from an understanding of the human’s reasoning structure.

Another line of work explores changing \textit{how} AI information is framed and presented~\cite{haselager2024, fischer2025,madumal2018, talk2model, zavolokina2024think}. 
It was shown that phrasing  explanations as reflective questions (e.g., ``\textit{If [reason], does it follow that [decision]?}") instead of declarative statements can improve humans' ability to detect logically flawed claims~\cite{danry2023}. Building on this insight, the rapid expansion of large language models (LLMs) in education and problem solving contexts~\cite{liu2024effects, chang2025generative, hui2025incorporating, yuan2024generative, educsci15081006} has made 
the ``\textit{Socratic LLMs}''~\cite{al2024can,hung2024, liu2024socraticlm, ding2024boosting}---systems that guide humans through step-by-step reasoning by posing scaffolded questions---a particularly prominent format for fostering critical reflection. These Socratic interactions now appear in major commercial products such as OpenAI's study mode \cite{openai}, Gemini's Guided Learning mode \cite{gemini}, and Claude for Education~\cite{claude}. However, whether in question-framed explanations or Socratic LLMs, the questions themselves typically follow the AI's own reasoning trajectory (with  Socratic deliberation~\cite{khadar2025wisdom} being an exception; discussed below). As such, the AI effectively steers humans 
along an AI-chosen line of inquiry and invites humans to reason \textit{with} it, rather than meeting humans where they are by adopting their (even incorrect) assumptions, reconstructing and diagnosing their argument, and systematically helping them recover from various missteps in their own reasoning.

Most closely related to our work are a few very recent frameworks that explicitly elicit people's internal reasoning processes. ``Human-AI Deliberation''~\cite{ma2025} is one such framework in which humans and AI engage in in-depth discussions about disagreements in their decisions and rationales. However, the elicited human reasoning is used primarily for surfacing divergences rather than diagnosing or correcting flaws within the reasoning itself, leaving humans responsible for interpreting these discrepancies (e.g., by digesting AI's justification) without structured support for metacognitive critique.
On the other hand, the ExtendAI framework proposed in~\cite{reicherts2025ai} collects natural-language decision rationales from humans and provides LLM-generated feedback, 
while the Socratic deliberation method proposed in~\cite{khadar2025wisdom} also elicits humans' rationale and engages them in reflective dialogues. 
Yet, in both cases, the feedback or reflective dialogue is produced entirely through LLM prompting, with no \textit{computational representation} of the human's reasoning structure or \textit{domain-grounded criteria} for rigorously evaluating its quality. As a result, the feedback often remains broad and generic, drawing on the LLM's general world knowledge rather than on a structured, quantitative  assessment of the completeness, reliability, or internal consistency of humans' reasoning grounded in domain-specific knowledge.

Table~\ref{tab:CTmethod} provides a comparison of existing reflection-oriented approaches in human-AI decision making. Across all approaches, reflection is primarily organized around AI outputs---timing, content, framing, or disagreements between AI and human decision-makers. Human reasoning either remains a black box or, when elicited, is used only for comparison or high-level feedback rather than for domain-grounded, structured analysis. AACT fills this gap by evaluating arguments from \textit{within the human's own perspective}---it elicits the human’s reasoning and models it computationally, adopts the human's perspective, and leverages the domain-specific model's reliable knowledge to perform counterfactual analysis over the human’s own argument and reveal where it breaks down. In doing so, AACT goes beyond telling humans \textit{how the AI thinks} but focuses on providing context-specific, quantitative guidance on analyzing \textit{how their own reasoning might fail}, shifting the role of AI from an information provider to a ``\textit{thought partner}'' \cite{collins2024}.

\subsection{Critical Thinking Methods and Frameworks}
Definitions of critical thinking include purposeful, self-regulatory judgment~\citep{abrami2008} and the ability to analyze and evaluate arguments \citep{lai2011}.  
To design AI assistance that promotes people's critical reflection of their own reasoning, we leverage established methodology and theories of critical thinking in education and clinical diagnosis.
Developing students' critical thinking skills is a long-standing goal in education \citep{bailin2002, behar2011}, where methods include Socratic questioning, which probes gaps in knowledge systematically \citep{SocraticQ}, and structured dialogue or debate, which encourages reflection and argumentation \citep{abrami2015}. 
On the other hand, in clinical diagnosis, extensive research studies the use of cognitive reasoning tools to improve clinical reasoning and reduce diagnostic errors \cite{staal2022}. Such techniques include checklists aimed to de-bias the diagnosis \cite{Taro2013, o2019cognitive} or encourage reflection \cite{kilian2019}, and guided reflection, which prompts clinicians to consider supporting, opposing, and missing evidence for multiple hypotheses \citep{Mamede2007, Mamede2008, Mamede2010, ilgen2013comparing, lambe2018guided, costa2019effects, staal2022, mamede2023deliberate}. 

The Recognition/Metacognition (R/M) model \citep{cohen1996, cohen1998critical} is particularly relevant to our work, and is often used in time-stressed, real-world tactical decision making. 
It describes decision making as a two-stage process: rapid recognition through pattern matching, followed by critiquing and correcting potential issues when warranted. 
Building on the foundations and history of critical thinking research, our AI-Assisted Critical Thinking framework adapts principles of the R/M model, but also leverages questioning \cite{SocraticQ} and structured dialogue \cite{abrami2015} to further develop critical thinking.

%% file: tables/background.tex
\begin{table*}[t!]
\centering
\small
\begin{tabular}{p{2.6cm} p{2.9cm} p{1.8cm} p{2.8cm} p{3.6cm}}
\toprule
\textbf{Approach} & \textbf{Core Mechanism} & \textbf{Elicit humans' \textit{own} rationale?} & \textbf{Reveal AI's decision or rationale?} & \textbf{Primary object of reflection} \\
\midrule

\textbf{Cognitive Forcing Functions~\cite{cff}} &
Change when or how AI advice is revealed & 
No &
Yes (delayed or partially) & AI's advice vs. human's (hidden) thinking\\
\midrule

\textbf{Hypothesis-Driven XAI~\cite{Miller2023}} &
Present evidence for and against multiple hypotheses &
No &
No (AI decision withheld; multiple rationales shown) & Multiple AI hypotheses and supporting/opposing evidence\\
\midrule

\textbf{Devil's Advocate~\cite{chiang2024}} &
Provide counter-evidence that opposes the human’s initial decision &
No &
No (AI is set to oppose;  opposing rationale shown) & AI's opposition vs. human's initial decision\\
\midrule

\textbf{Reflective \newline Questioning~\cite{danry2023}} &
Frame AI explanations as questions  &
No (only reasoning to {\em AI's questions} elicited) &
No & Reflective questions exposing an AI-selected causal link ``[reason]$\rightarrow$[decision]''\\
\midrule

\textbf{Socratic LLMs in \newline Education (e.g., \cite{liu2024socraticlm})} &
Multi-turn guided questioning to scaffold problem solving &
No (only reasoning to {\em AI's questions} elicited)
&
Maybe (incrementally revealed along with interactions with humans) & AI-generated tutoring questions (often reflect AI's reasoning path)\\
\midrule
\textbf{Socratic LLMs for \newline Deliberation \cite{khadar2025wisdom}} &
LLMs act as deliberation partner to interrogate the human's hypothesis&
Yes
&
No & Human's rationale viewed through AI's questions (interrogative prompts not grounded in domain-specific knowledge)\\
\midrule
\textbf{Human--AI \newline Deliberation~\cite{ma2025}} &
Dialogue about disagreement in predictions/rationales &
Yes &
Yes & AI's decision/rationale vs. human's decision/rationale\\
\midrule

\textbf{ExtendAI~\cite{reicherts2025ai}} &
Provide LLM-generated feedback to human decision rationale &
Yes &
No & Human's rationale viewed through AI's evaluative lens (high-level feedback not grounded in domain-specific knowledge)\\
\midrule

\textbf{AACT (this work)} &
Counterfactual analysis of the human’s decision argument and structured metacognitive support &
Yes &
No & Human's own decision argument (potential incompleteness, unreliability, conflict issues, grounded in domain-specific model's counterfactual analysis)\\
\bottomrule
\end{tabular}
\vspace{2pt}
\caption{A comparison of reflection-oriented approaches in human--AI decision making.}~\label{tab:CTmethod}
\vspace{-10pt}
\end{table*}

%% file: sections/3_system.tex
\section{AI-Assisted Critical Thinking: Framework and Instantiation}
\label{sec:system}

In this section, we introduce  the AI-Assisted Critical Thinking (\aact{}) framework. 
We start by providing an overview of the ``{\em Recognition/Metacognition (R/M) model}''~\cite{cohen1996,cohen1998critical}, which formally describes the deliberate metacognitive processes and associated critical thinking strategies. Then, we present the \aact{} framework, which builds upon the R/M model and adapts the critique and correction components by leveraging a domain-specific AI model to adopt the human decision-maker's perspective and conduct counterfactual analysis of their reasoning, thereby supporting the critique and correction of it. Finally, we discuss the design space underlying the \aact{} framework and detail our instantiation of \aact{} as a conversational AI system to facilitate human decision-makers' critical thinking during decision making.

\subsection{Theoretical Background: The Recognition/Metacognition Model}
\label{sec:RM}

The Recognition/Metacognition (R/M) model \cite{cohen1996,cohen1998critical} is a framework that provides structured critical thinking strategies for decision making under time pressure and uncertainty.
In this model, ``recognition'' refers to quick and intuitive decision making based on pattern-based recognition. 
In contrast, ``metacognition'' 
is the understanding and analysis of one's own decisions, such as critiquing and correcting recognition-based judgments. More formally, the R/M model includes four essential components: 
\begin{itemize}
\item \textbf{Recognize}: Corresponding to ``recognition'' stage of the R/M model, given a decision making task, the decision-maker first identifies possible evidence-conclusion relationships for the task---that is, the patterns they use to map observed cues to a decision (e.g., ``when cue $A$ is observed, decision $B$ can be inferred''), which constitute their decision ``{\em arguments}''.

\item \textbf{Critique}: As part of the ``metacognition'' stage, given a decision and the arguments associated with it, 
the critiquing process aims to identify problems within those  arguments. 
Three types of problems can be identified as a result of the critiquing process: (1) {\em Incompleteness}: An argument is incomplete if there exists additional information that has not been considered in the argument that may either confirm or disconfirm their conclusion;  (2) {\em Unreliability}:  An argument is unreliable if the link between evidence and conclusion is based on unexamined or doubtful assumptions; and 
(3) {\em Conflict}: Conflicts in arguments occur when beyond the current argument, there exist alternative arguments that provide support for conflicting conclusions. 

\item \textbf{Correct}: Following the critique of decision arguments, another metacognitive step involves decision-maker responding to the problems identified. These corrections often involve decision-makers retrieving more information, both internally and externally, and revisiting and revising their assumptions.  In doing so, decision-makers fill gaps in their judgment, strengthen the reliability of their arguments, resolve conflict among competing considerations, and eventually formulate a single coherent, and potentially more comprehensive argument. 

\item \textbf{Quick-test}: This is a cost-benefit analysis that regulates the initiation and duration of metacognition (i.e., critique and correct), accounting for factors such as time pressure, the uncertainty/novelty of the decision scenario, and the stakes of the decision. For example, in high-stake tasks with uncertainty, quick-test may deem metacognition and critical thinking to be beneficial, while in time-stressed settings where the decision-maker has high confidence, metacognition may not be necessary and decisions can be made based on immediate recognitional response.

\end{itemize}

The R/M model not only decomposes the metacognitive processes of critique and correction, but also provides a structured categorization of reasoning flaws and correction strategies. This makes it a well-suited foundation for AACT, where we support and operationalize the metacognitives processes with AI assistance. Moreover, the R/M model was developed for time-stressed decision making scenarios with uncertainty, incomplete information, and high-stake outcomes, and see application in tactical decision making and training. These conditions closely mirror 
many human-AI decision making contexts, such as medical diagnosis or judicial judgments, 
where people often need to rapidly interpret information, integrate cues, manage uncertainty, and avoid cognitive biases.

\subsection{The AI-Assisted Critical Thinking Framework}
\label{sec:framework}

\begin{figure*}[t!]
    \center
    \includegraphics[width=\textwidth]{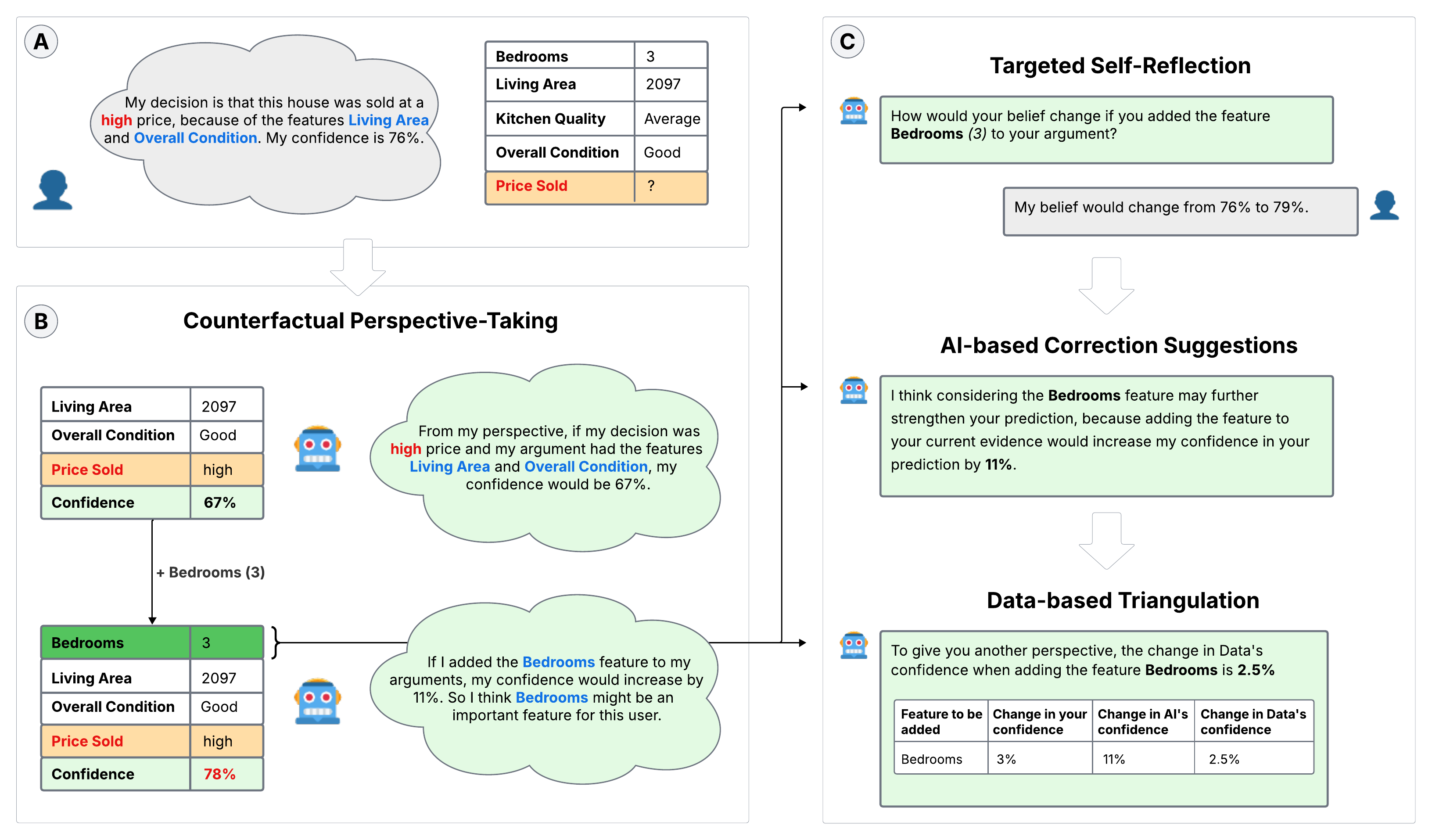}
    \caption{
    Illustrative example of how \aact{} supports decision-makers in critiquing and correcting their argument. (A) The decision-maker makes a decision, selects some features as their argument, and reports their confidence. (B) The domain-specific AI model performs \textit{counterfactual perspective-taking} by adopting the decision-maker's perspective to evaluate its confidence in the decision-maker's decision, then conducting counterfactual analysis by adding or removing features from their argument. (C) Using AI's counterfactual perspective-taking results, \aact{} prompts the decision-maker to critique their argument through \textit{targeted self-reflection}, supports revisions with 
    \textit{AI-based correction suggestions}, and 
    enables 
    \textit{data-based triangulation}.
    }
    \label{fig:critique}
\end{figure*}

Inspired by the R/M model, we present the AI-Assisted Critical Thinking (\aact{}) framework which explicitly designs AI to facilitate critical thinking and metacognition---that is, supporting decision-makers in critiquing and correcting their own decisions and arguments (see Figure~\ref{fig:critique} for an illustrative overview).

The key mechanism through which \aact{} supports critique and correction of a decision-maker's internal reasoning is by engaging an AI model in \textit{counterfactual perspective-taking} (Section~\ref{sec:cpt}). This involves two steps: (1) \textit{role playing}, in which the AI model ``adopts the decision-maker's perspective'' by taking their argument and asking ``\textit{If I follow the human's argument, how confident am I in the human's decision?}'', and (2) \textit{counterfactual analysis}, in which the AI perturbs elements of the human's argument and asks ``what-if'' questions---``\textit{If I change this part of the human's argument, how would that change my confidence in the human's decision or an alternative decision?}'' 
Leveraging counterfactual perspective-taking, the AI model can identify potential issues of incompleteness, unreliability, and conflicts in the human's argument,
and then enables
decision-makers to 
\textit{critique} these identified issues by themselves via counterfactual questioning
(Section~\ref{sec:critique}). 
To further help human decision-makers \textit{correct} these potential issues, the AI provides its own suggestions and supplementary insights drawn from its training data (Section~\ref{sec:correct}).

Without loss of generality, we assume that a decision making task can be characterized by $N$ features $(X_1,\ldots,X_N)$. 
For a specific decision making task $\boldsymbol{x}=(X_1=v_1, X_2=v_2, \ldots, X_N=v_N)$ with a ground truth of $y\in \mathcal{Y}=\{Y_1,\ldots,Y_C\}$, we denote the decision-maker's  decision as $y_h$ ($h$ for human), which reflects their recognition-based judgment of the task. To elicit the decision-maker's internal recognitional reasoning process, in AACT, we first ask 
the decision-maker to select a subset of features in the task as their \textit{argument}, formalized as $\boldsymbol{x_h}\subseteq \{X_1=v_1,X_2=v_2,\ldots,X_N=v_N\}$. 
For example, $\boldsymbol{x_h}=\{X_2=v_2, X_N=v_N\}$ means that the decision-maker concludes the decision is $y_h$ because of the features $X_2$ and $X_N$ in task $\boldsymbol{x}$. 
In addition, we also ask the decision-maker to rate their {\em confidence} in their decision, denoted as $p_h=P_H(y=y_h|\boldsymbol{x_h})$\footnote{The subscript in the probability function $P(\cdot)$ is used to differentiate whether the probability is computed by the human decision-maker (``H''), the AI model (``M''), or based on a dataset (``D'').}---that is, the decision-maker's belief about the likelihood of their decision $y_h$ being correct based on their argument $\boldsymbol{x_h}$. Finally, to support decision-maker's metacognitive processes, i.e., the critique and correction of their recognition-based decision, the \aact{} framework leverages a domain-specific AI model $m$ that is trained on a dataset of $S$ feature-decision pairs $D=\{(\boldsymbol{x_i}, y_i)\}|_{i=1}^{S}$. 
We constrain the AI model to be a {\em probabilistic classifier}, i.e., it can predict the probability $P_M(y=y'|\boldsymbol{x})$ for an input $\boldsymbol{x}$ belonging to any class $y'\in\mathcal{Y}$  
(see Table ~\ref{table:notations} for a summary of key notations in AACT).

\subsubsection{Engaging AI for Counterfactual Perspective-Taking}
\label{sec:cpt}

To enable the AI model to perform counterfactual perspective-taking, we leverage its probability function for \textit{role playing} and \textit{counterfactual analysis}. 
In role playing, the AI adopts human's perspective by computing $p_m=P_M(y=y_{h}|\boldsymbol{x_{h}})$---its confidence in the human's decision given the human's argument.
Then, the AI conducts counterfactual analysis by evaluating perturbed arguments $\boldsymbol{x_{h}'}$ (derived from the human's original argument $\boldsymbol{x_{h}}$) and computing $P_M(y=y^{*}|\boldsymbol{x_{h}^{'}})$ for any possible decision $y^{*}$.  These computed probabilities will then inform the critique and correction of the human's argument. 
 
To compute $P_M(y=y_{h}|\boldsymbol{x_{h}})$ and $P_M(y=y^{*}|\boldsymbol{x_{h}^{'}})$, we effectively need to compute AI's confidence on any decision $y^{*}\in\mathcal{Y}$ given any argument $\boldsymbol{x^{*}} \subseteq \boldsymbol{x}$, i.e., $P_M(y=y^{*}|\boldsymbol{x^{*}})$---or for brevity,  $P_M(y^{*}|\boldsymbol{x^{*}})$. 
Although the AI model directly provides $P_M(y^{*}|\boldsymbol{x})$ only for the full feature set  $\boldsymbol{x}$,  the quantity $P_M(y^{*}|\boldsymbol{x^{*}})$ can be obtained by marginalizing over the features not included in $\boldsymbol{x^*}$, which we approximate through Monte Carlo sampling (computation details can be found in Appendix~\ref{app:math1}).

\subsubsection{\aact{} Supports for Critiquing Arguments}
\label{sec:critique}

According to the R/M model, the goal of the critiquing process is to identify incompleteness, unreliability, and conflict problems within the decision-maker's argument.  
Below, we describe how the AI model  identify these potential 
issues 
in the human decision-maker's argument using its counterfactual perspective-taking results, and how these findings are then used to prompt decision-makers to reflect on these issues through targeted counterfactual questioning.

\vspace{2pt}
\noindent \textbf{AI's identification of incomplete argument.} 
The AI model can identify incompleteness issues in arguments by identifying features overlooked by the decision-maker that could significantly influence its confidence in the human's decision $y_h$.  
Specifically, for each feature $X_j$ that is not included in the decision-maker's argument (i.e., $X_j\notin \boldsymbol{x_h}$), the AI model computes  
$\Delta p_m^{\boldsymbol{x_h}+X_j} := P_M(y_h|\boldsymbol{x_h}\cup \{X_j=v_j\}) - P_M(y_h|\boldsymbol{x_h})$, which is the change of AI's confidence in human's decision when including feature $X_j$ in human's argument. We define a threshold parameter $\epsilon\in[0,1]$ to determine if a change is ``substantial,''
 and we note two cases of interests:
\begin{itemize}
\item When $\Delta p_m^{\boldsymbol{x_h}+X_j}>\epsilon$, it means that AI finds including the overlooked feature $X_j$ into the argument would significantly {\em increase} its confidence in the human's decision, thus $X_j$ can be considered as a \textit{missing supporting feature} that could be used to further strengthen the decision-maker's argument. 

\item In contrast, when $\Delta p_m^{\boldsymbol{x_h}+X_j}<-\epsilon$, it means that AI finds including the overlooked feature $X_j$ into the argument would significantly {\em decrease} its confidence in the human's decision, thus $X_j$ can be considered as a \textit{missing opposing feature} that might weaken the decision-maker's argument. 
\end{itemize}

In both cases, the human decision-maker's own reasoning process may suffer from incompleteness issues.

\input{tables/notations}

\vspace{2pt}
\noindent \textbf{AI's identification of unreliable argument.} 
The AI model highlights reliability issues in arguments by identifying some features included in the decision-maker's argument that may not sufficiently support their decision $y_h$. 
Specifically, for each feature $X_j$ that is included in the decision-maker's argument (i.e., $X_j\in \boldsymbol{x_h}$), the AI model computes  
$\Delta p_m^{\boldsymbol{x_h}-X_j} := P_M(y_h|\boldsymbol{x_h} - \{X_j=v_j\}) - P_M(y_h|\boldsymbol{x_h})$---the change of AI's confidence in human's decision when excluding feature $X_j$ and its value from the decision argument. 
Similar as that in identifying incomplete arguments, we can characterize the type of feature $X_j$ by comparing $\Delta p_m^{\boldsymbol{x_h}-X_j}$ to the threshold $\epsilon$.
$\Delta p_m^{\boldsymbol{x_h}-X_j}>\epsilon$ means removing feature $X_j$ from the argument \textit{increases} AI's confidence in the human's decision; in this case, $X_j$ can be considered as an \textit{unreliable feature} that may counteract the decision-maker's conclusion. 
However, if $\Delta p_m^{\boldsymbol{x_h}-X_j}<-\epsilon$, then we can infer that AI finds $X_j$ to be an important feature for supporting the decision $y_h$. That is, AI agrees with the decision-maker that it is a \textit{reliable feature}. Finally, when $|\Delta p_m^{\boldsymbol{x_h}-X_j}|\le\epsilon$, it means removing feature $X_j$ from the argument does not significantly change AI's confidence in the human's decision; in this case, $X_j$ can be considered as an \textit{irrelevant feature} since the link between it and the conclusion is weak.

If there exists some feature in the human's argument that is identified as unreliable or irrelevant, it implies that the human's own reasoning may suffer from unreliability issues.

\vspace{2pt}
\noindent \textbf{AI's identification of conflicts in arguments.}
The AI model can also surface potential conflicting arguments by identifying new arguments different from the decision-maker's argument that provide strong support for an alternative decision $y_c$ that is different from the human's decision $y_h$. 

To generate strong conflicting arguments, for each alternative decision $y_c\neq y_h$, AI can identify the ``strongest argument'' for that decision $\boldsymbol{x_c}= \arg\max_{\boldsymbol{x_c^{'}}\subseteq\boldsymbol{x}} P_M(y_c|\boldsymbol{x_c^{'}})$, which is a subset of features such that when focusing only on their values, 
AI's confidence in $y_c$ reaches the highest. In practice, searching through all possible feature subsets to identify $\boldsymbol{x_c}$ can be computationally challenging,
so we approximate the computation of $\boldsymbol{x_c}$ using post-hoc explainable AI methods (see Appendix~\ref{app:math2} for details).

With the argument identified for each alternative decision, AI can rank them by $p_m^c= P_M(y_c|\boldsymbol{x_c})$---its confidence in the alterative decision given its strongest argument---and 
compare the confidence with the chance of a random guess (i.e., $\frac{1}{C}$). If some alternative decision and its associated argument satisfy $p_m^c>\frac{1}{C}$, it means 
the AI's confidence in the alternative decision $y_c$ given its argument is higher than taking a random guess, which may suggest the human's own reasoning has conflict issues.

\input{tables/FS}

\vspace{2pt}
\noindent \textbf{Supporting decision-makers' critique of their own argument via targeted self-reflection.} 
Table~\ref{table:FS} provides a summary of how the AI model can use its counterfactual perspective-taking results to identify different types of issues in the decision-maker's argument. However, the essence of ``critiquing'' is  enabling decision-makers to recognize these issues \textit{themselves} through critical reflection. 
To support this, we introduce a ``\textit{targeted self-reflection}'' mechanism that uses counterfactual questioning to prompt decision-makers to consider how their decision might change under specific hypothetical modifications to their argument. These hypothetical changes are selected based on potential issues identified by the AI model, directing the decision-maker's attention to aspects of their reasoning that may require more scrutiny. More specifically:

\begin{itemize}
\item For incompleteness issues, when AI identifies $X_j$ as a missing supporting feature or missing opposing feature, the decision-maker is asked to reflect how adding feature $X_j$ to their argument would change their own confidence in their decision (i.e., to evaluate $\Delta p_h^{\boldsymbol{x_h}+X_j}:= P_H(y_h|\boldsymbol{x_h} + \{X_j=v_j\}) - P_H(y_h|\boldsymbol{x_h})$).
\item For unreliability issues, when AI identifies $X_j$ as an unreliable feature or irrelevant feature, the decision-maker is asked to reflect how removing feature $X_j$ from their argument would change their own confidence in their decision (i.e., to evaluate $\Delta p_h^{\boldsymbol{x_h}-X_j}:= P_H(y_h|\boldsymbol{x_h} - \{X_j=v_j\}) - P_H(y_h|\boldsymbol{x_h})$).
\item For conflict issues,
for each of the top $k$ alternative decision $y_c$ that the AI model is most confident about (i.e., with the highest $p_m^c$), 
the decision-maker is asked to estimate their own confidence in the alternative decision $y_c$ when focusing only on features in $\boldsymbol{x_c}$ (i.e., to evaluate $p_h^c=P_H(y_c|\boldsymbol{x_c})$) ($k$ is a system parameter controlling the number of alternative decisions for the decision-maker to reflect on). 
\end{itemize}

\noindent See Figure~\ref{fig:ui_aact1} for an example user interface for targeted self-reflection on incompleteness issues.

Via counterfactual questioning, the decision-maker is prompted to closely examine specific parts of their argument and explicitly articulate their underlying beliefs, which is an approach shown to improve critical thinking~\cite{chi1994eliciting}. 
Crucially, targeted self-reflection is elicited \textit{before} the presentation of the AI's suggestions on argument correction (see details in Section~\ref{sec:correct}), 
ensuring that decision-makers are not biased by the AI's counterfactual analysis. 
This design also implicitly stimulates the decision-maker's curiosity about the AI's suggestions by making gaps in their knowledge more salient~\cite{loewenstein1994psychology}, a strategy known to improve information comprehension~\cite{kim2017explaining}.

\subsubsection{\aact{} Supports for Correcting Arguments}
\label{sec:correct}

After critiquing their decision argument, decision-makers then need to correct the identified issues in their argument by revisiting and revising their assumptions, which is typically enabled by retrieving  additional information either internally from their long-term memory (e.g., their domain knowledge) or externally through further data and observation. In \aact{}, we support this information retrieval process through two mechanisms: \textit{AI-based correction suggestions}, which explain to decision-makers how and why AI recommends revising the argument, and \textit{data-based triangulation}, which provides decision-makers with insights drawn from empirical patterns of the AI model's training dataset as an additional source of information. 
See Figure~\ref{fig:ui_aact2} for an example user interface for AI-based correction suggestions and Figure~\ref{fig:ui_aact3} for data-based triangulation.

\vspace{2pt}

\noindent \textbf{AI-based correction suggestions.} Recall that each hypothetical argument change presented to decision-makers during targeted self-reflection is selected because the AI model associates that change with a potential issue in the decision-maker's argument. In other words, each hypothetical change corresponds to a possible correction of the argument identified by the AI model. Thus, to support the correction of decision argument, after decision-makers reflect on a given  hypothetical change, \aact{} directly presents them with AI's ``correction suggestions'' that explains both \textit{what} revision the AI recommends and \textit{why} it is needed. For example, when the hypothetical change involves a missing supporting feature $X_j$, the AI can justify its suggestion of including $X_j$ into the argument by stating that ``adding feature $X_j$ to the argument will boost the confidence in the current decision by $\Delta p_m^{\boldsymbol{x_h}+X_j}$''. Conversely, if the hypothetical change reveals an unreliable feature $X_j$, the AI can suggest removing it and explain that ``excluding feature $X_j$ to the argument will actually increase the confidence in the current decision by $\Delta p_m^{\boldsymbol{x_h}-X_j}$''. Through these explanations, the AI provides decision-makers with additional information---derived directly from its counterfactual analysis---about dependencies among features and their impact on decision confidence, thereby supporting the information retrieval process needed to correct their argument.

\vspace{2pt}
\noindent \textbf{Data-based triangulation.} 

After completing self-reflection and reviewing the AI's correction suggestions, 
decision-makers need to assess the trustworthiness of AI suggestions and decide how to act upon them. In practice, it is common for decision-makers and AI models to hold different beliefs (e.g., the decision-maker's estimate of how adding $X_j$ would affect confidence in the current decision---$\Delta p_h^{\boldsymbol{x_h}+X_j}$---may differ substantially from AI's estimate, $\Delta p_m^{\boldsymbol{x_h}+X_j}$). Reconciling this difference often requires additional information. In \aact{}, we support this reconciliation through data-based triangulation, which provides 
empirical insights derived from 
the AI model's training dataset. 
Specifically, for each probability $P_M(y^{*}|\boldsymbol{x^{*}})$ computed by the AI model, we compute its empirical counterpart $P_D(y^{*}|\boldsymbol{x^{*}})$ by (1) selecting training data instances whose features match the values in $\boldsymbol{x^*}$ and (2) counting the fraction of those instances whose ground truth decision is $y^*$.
Using these empirical probabilities, we can then compute
 $\Delta p_d^{\boldsymbol{x_h}+X_j}
 $ for incompleteness issues, 
$\Delta p_d^{\boldsymbol{x_h}-X_j}$ for unreliability issues, and $p_d^c$ for conflict issues\footnote{ 
In some cases, there may exist very few data instances in a dataset matching the feature values specified in $\boldsymbol{x^*}$, so the empirical probabilities generated may not be meaningful. In this case, the system could report the corresponding data probabilities as ``not available.''}.

Therefore, for each correction suggestion provided by the AI model, the decision-maker can triangulate among their own beliefs (e.g., $\Delta p_h^{\boldsymbol{x_h}+X_j}$), the AI’s suggestions (e.g., $\Delta p_m^{\boldsymbol{x_h}+X_j}$), and the empirical data distributions (e.g., $\Delta p_d^{\boldsymbol{x_h}+X_j}$), thereby forming a richer basis for assessing trustworthiness of the AI's suggestions and decide how to act upon them.
For instance, a decision-maker may regard an AI suggestion to include a feature in their argument as trustworthy when $\Delta p_m^{\boldsymbol{x_h}+X_j}$ aligns more closely with $\Delta p_d^{\boldsymbol{x_h}+X_j}$ than their own $\Delta p_h^{\boldsymbol{x_h}+X_j}$ does. Conversely, if the empirical data distribution appears counterintuitive, the decision-maker may infer that the AI model's training dataset is biased and withhold trust in AI suggestions, even when they match the empirical distribution well. 
Ultimately, how decision-makers interpret and act on this information is at their discretion and lies beyond the scope of the \aact{} framework.

\subsection{Design Space of the \aact{} Framework}
\label{sec:dc}

In this section, we elaborate the design space underlying the operationalization of \aact{} into a functional AI system. Specific choices along different dimensions of this design space are highly context-dependent and may vary with factors such as the decision-maker's efficiency in processing information, tolerance for cognitive load, time constraints of the decision task, and decision stakes. Further discussions on how these choices may differ across applications, informed by our user study findings, are provided in Section~\ref{sec:design}.

\subsubsection{Workflow of Critique and Correction} 

As described in Section~\ref{sec:framework}, for a specific issue identified by the AI model, the decision-maker generally proceeds through a structured, three-step critique-and-correction workflow of 
``targeted self-reflection
$\rightarrow$
AI-based correction suggestion
$\rightarrow$
data-based triangulation''
(see Figure~\ref{fig:critique}C for an illustrative example). 
However, when implementing \aact{}, several higher-level workflow design choices must still be specified.

The first decision is the \textit{scope of critique}: whether the AI should focus exclusively on problematic aspects of the decision-maker's argument and guide them through the critique-and-correction workflow for all issues it identifies, or also highlights point of agreement, such as those ``reliable features'' the AI considers to be sound. 

A second design choice concerns the \textit{temporal structure} in presenting various issues. When the AI model detects multiple issues in the decision-maker's argument, they may be presented synchronously, prompting the decision-maker to consider all problems holistically, or sequentially, requiring the decision-maker to complete the critique-and-correction workflow for each issue one at a time. 
If the issues are presented sequentially, a third design choice is \textit{the order of presentation}. One option is an ``agreement-to-disagreement order'' (e.g., 
``agreement
$\rightarrow$
incompleteness
$\rightarrow$
unreliability
$\rightarrow$
conflict'')
: the AI first highlights the reliable features it shares with the decision-maker, then surfaces the incompleteness and unreliability issues with the aim of strengthening the decision-maker's existing argument, and finally introduces alternative hypotheses that challenge and potentially disrupt the decision-maker's reasoning. This approach aligns with the principles of Rogerian argumentation~\cite{rogers1995becoming,young1970rhetoric}, which recommends acknowledging common ground first and gradually introducing areas of divergence in argumentation. Another approach is to adopt an adaptive order based on AI's confidence in the alternative decision (e.g., $P_M(y_c|\boldsymbol{x})$ or $P_M(y_c|\boldsymbol{x_c})$)---for instance, AI could prioritize presenting conflicting arguments if its confidence in the alterative decision is high and prioritize highlighting agreement otherwise.

Another decision consideration in the \aact{} workflow concerns \textit{when and how decision-makers may update their argument or decision}. 
The system may restrict updates until all issues have been presented, or it may allow granular updates, enabling decision-makers to revise their argument after each issue is considered. In the latter case, a subsequent design choice is how to proceed after each update. Two natural options are to (1) ``\textit{continue},'' in which the system re-identify and present issues for the revised argument but limits this to issue types that have not yet been covered, or to (2) ``\textit{restart},'' requiring the system to fully re-analyze the revised argument and present all types of issues again from the beginning.

\subsubsection{System Parameters} 
\aact{} involves several tunable parameters, such as the threshold determining if a change in AI's confidence is significant (i.e., $\epsilon\in[0,1]$) and the max number of alternative decisions decision-makers are prompted to consider corresponding to the AI-identified conflicts in argument (i.e., $k$). 
Another implicit parameter is \texttt{max\_feature\_change}, the max number of features that can be simultaneously added to or removed from the decision-maker's argument in the counterfactual analysis (in Section~\ref{sec:critique}, we illustrate the case where \texttt{max\_feature\_change}$=1$). 

Setting these parameters to different values will impact the decision-maker's interaction experience with an \aact{} system. For example, a higher $\epsilon$ is generally associated with the system identifying fewer issues for the decision-maker's argument, while a lower $\epsilon$ is often associated with more detected issues. In addition, larger $k$ values will lead to the evaluation of more alterative decisions, resulting in a more comprehensive critique of potential conflict issues in one's argument.
Finally, increasing the value of \texttt{max\_feature\_change} allows the system to account for more complex interactions between features in the counterfactual analysis, but might also increase the difficulty for decision-makers to understand its results.

\subsubsection{Supports for quick-test}
\label{sec:qt}
In Section~\ref{sec:framework}, we describe how \aact{} can facilitate the critiquing and correcting processes---the two metacognitive components in the R/M model---in decision making. Another component in the R/M model that involves higher-order thinking is the ``quick-test'', which determines whether the metacognitive critical thinking process should be initiated and for how long it should continue. Deciding whether and how to support quick-test is another design choice in \aact{}. 
One option is to leave quick-test entirely to the decision-maker, allowing them to request for AI supports in critiquing and correcting arguments on demand only when they deem necessary. Alternatively, the system may estimate the ``seriousness'' of each issue the AI model identifies in the decision-maker's argument, and automatically trigger the critique-and-correction workflow for issues that are most consequential. 
A most sophisticated option is to perform a utility analysis that considers factors like decision stakes, time constraints, and the confidence of both the decision-maker and the AI, 
and initiate metacognitive processing only when the expected benefit of doing so outweighs its cost.

\begin{figure*}[t!]
    \center
    \includegraphics[width=\textwidth]{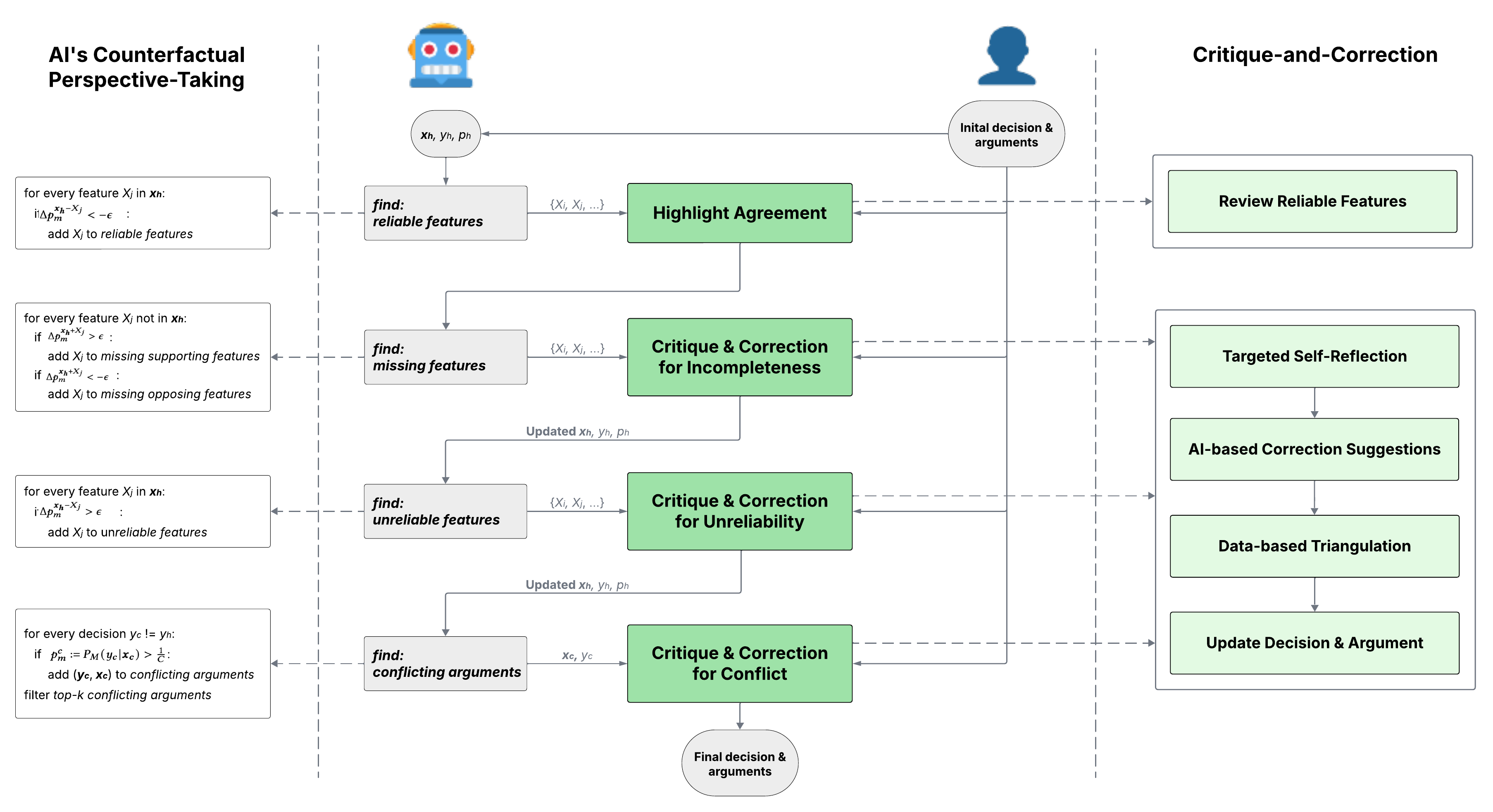}
    \caption{Flowchart of our \aact{} implementation. 
    (1) Middle panel: the main workflow of presenting the three types of issues in decision-maker's argument, along with highlighting agreement  between AI and human on reliable features. 
    (2) Left panel: the issues and their associated features are identified using AI's counterfactual perspective-taking results.
    (3) Right panel: For the incompleteness, unreliability, and conflict stages, \aact{} engages decision-makers in the critique-and-correction workflow; at the end of each stage, decision-makers can update their decision and argument.}
    \label{fig:flowchart}
\end{figure*}

\subsection{Instantiating the \aact{} Framework as a Conversational AI System}
\label{sec:implement}

Inspired by recent work leveraging dialogue to improve human-AI interaction~\cite{madumal2018, talk2model} and promote critical thinking~\cite{abrami2015,ma2025,khadar2025wisdom}, we instantiate the \aact{} framework as a conversational AI system. 
Each component of the critique-and-correction workflow (i.e., targeted self-reflection, AI-based correction suggestions, data-based triangulation) is delivered to decision-makers as chat messages that are composed of
fixed, pre-defined sentence templates with slots dynamically filled with the AI model's counterfactual perspective-taking results (e.g., feature names, change in confidence)\footnote{The decision to use pre-defined templates to generate AI messages ensures that every message is consistent, unambiguous, and tightly coupled to the underlying counterfactual analysis. Compared to using LLMs to generate the conversational messages, this approach avoids hallucinations and eliminates linguistic variability and stylistic difference in the messages received by different decision-makers, thereby isolating the behavioral effects of the AACT mechanisms themselves rather than the writing quality or conversational style of an LLM. Nonetheless, we discuss in Section~\ref{sec:additionaldesign} how future instantiations may incorporate LLMs to provide richer, more adaptive conversational scaffolding.};
the full list of templates are provided in supplementary materials Section A.
The decision-maker can then respond to these messages to reflect and update their argument/decision through lightweight input mechanisms including sliders, radio buttons, and checkboxes. 

Figure~\ref{fig:flowchart} presents the flowchart for the instantiation of the \aact{} framework that we implemented for our user study. 
Given a decision making task, the system first collects the decision maker's decision $y_h$, argument $\boldsymbol{x_h}$, and decision confidence $p_h$. 
Then, the AI system identifies all issues in the decision-maker's argument using the criteria listed in Table~\ref{table:FS} and presents these issues to the decision-maker sequentially in the order of  ``agreement
$\rightarrow$
incompleteness
$\rightarrow$
unreliability
$\rightarrow$
conflict''. 
The initial ``agreement'' stage highlights the reliable features on which the AI and the decision-maker align; because no correction is needed, targeted self-reflection and data-based triangulation are omitted at this stage.
In the subsequent incompleteness, unreliability, and conflict stages, the AI engages the decision-maker in the full critique-and-correction workflow to address each type of issues in turn.
Within each of these three stages, the interaction between the AI system and the decision-maker unfolds following the 
``targeted self-reflection
$\rightarrow$
AI-based correction suggestion
$\rightarrow$
data-based triangulation'' 
workflow.  
Further, at the end of each stage, the decision-maker may update their decision, argument and confidence, and the AI's counterfactual analysis for later stages is based on these updated decision and argument.

More specifically, in the incompleteness stage,  
the AI system begins by prompting  
targeted self-reflection, asking  the decision-maker,  ``How would your confidence change if you added the feature $X_j$ to your argument?''  The decision-maker can respond using a slider (Figure ~\ref{fig:ui_aact1}).
Then, the AI provides its correction suggestion with a  statement like ``I think $X_j$ would strengthen/weaken your prediction, because adding the feature to your current evidence would increase/decrease my confidence in your prediction by $\Delta p_m^{\boldsymbol{x_h}+X_j}$ '' (Figure ~\ref{fig:ui_aact2}). 
To support triangulation, 
the system also presents
a side-by-side comparison showing 
how adding $X_j$ into the argument affects the decision-maker, AI, and data's confidence in the current decision (Figure ~\ref{fig:ui_aact3}). 
Finally, after reviewing all this information, the decision-maker is asked whether they would like to make any changes to their decision, arguments, or confidence, and they can respond using quick input options (Figure ~\ref{fig:ui_aact4}).

The unreliability and conflict stages follow a dialogue structure similar to that of the incompleteness stage. Specifically, for unreliability issues,
our early implementation included AI's correction suggestions on both unreliable and irrelevant features;
however, pilot user feedback indicated that this produced information overload. 
Accordingly, 
the final system used in our user study focuses only on 
critiques and corrections for unreliable features in the decision-maker's argument. 
Targeted self-reflection is presented as a question from AI ``How would your confidence change if you removed the feature $X_j$ from your argument?'', and the corresponding correction suggestion is provided via messages like ``I believe $X_j$ may not reliably support your prediction, because removing the feature from your current evidence would increase my confidence in your prediction by $\Delta p_m^{\boldsymbol{x_h}-X_j}$''.  
For conflict issues, 
targeted self-reflection is prompted with the AI question ``What would be your confidence in the alternative prediction $y_c$ if you only considered the features in $\boldsymbol{x_c}$?'', and the correction suggestion is provided by AI stating that  
``I think the alternative prediction $y_c$ might be possible when considering only the features in $\boldsymbol{x_c}$, as my confidence in $y_c$ is 
$p_m^c$ 
when focusing only on these features.'' 

Note that any of the four stages may be skipped if the AI's counterfactual analysis does not identify the corresponding issue.  
However, once a particular issue type is detected, the decision-maker is required to complete the full critique-and-correction workflow for that stage---that is, we do not apply the quick-test to
skip stages in our study, as our goal was to 
comprehensively evaluate all components of \aact{} in the user study.
Finally, we set the system parameters to $\epsilon=0.04$, $k=1$, 
and \texttt{max\_feature\_change}$=1$ to balance the generation of meaningful and easily-understandable critiques while controlling the number of them.

%% file: tables/notations.tex
\begin{table*}[t!]
\small
\begin{center}
\begin{tabular}{l p{3cm} p{8.5cm}}
\toprule
\textbf{Variable} & \textbf{Type} & \textbf{Description} 
\\
\midrule
$y_h$  & Decision  &   Human's decision              
\\[1.5mm]
$\boldsymbol{x_h}$   &  Argument  & Human's argument                
\\[1.5mm]
$p_h=P_H(y=y_h|\boldsymbol{x_h})$   & Confidence   & Human's confidence in their decision given their argument                 
\\[1.5mm]
$p_m=P_M(y=y_h|\boldsymbol{x_h})$    & Confidence   &  AI's confidence in human's decision given their argument               
\\[1.5mm]
$\Delta p_m^{\boldsymbol{x_h}+X_j}$      &   Change in Confidence & Change in AI's confidence in human's decision when including feature $X_j$ in human's argument               
\\[1.5mm]
$\Delta p_h^{\boldsymbol{x_h}-X_j}$      &   Change in Confidence & Change in AI's confidence in human's decision when excluding feature $X_j$ from human's argument                 
\\[1.5mm] 
$\epsilon$  &  System Parameter & Threshold, between 0 and 1, to determine if a change in AI's confidence is significant
\\[1.5mm]
$y_c$   &   Decision    &   Alternative decision $y_c\neq y_h$
\\[1.5mm]
$\boldsymbol{x_c}$  &   Argument    &   The strongest argument for the alternative decision $y_c$
\\[1.5mm]
$p_m^c=P_M(y=y_c|\boldsymbol{x_c})$    & Confidence   &  AI's confidence in the alternative decision $y_c$ given its strongest argument
\\[1.5mm]
$p_d^*=P_D(y=y^*|\boldsymbol{x^*})$    & Confidence   &  ``Data's confidence'' in a decision $y^*$ given an argument $\boldsymbol{x^*}$, used in computation for data-based triangulation.              
\\
\bottomrule
\end{tabular}
\end{center}
\vspace{2pt}
\caption{A summary of the key notations in \aact{}.} 
\label{table:notations}
\vspace{-15pt}
\end{table*}

%% file: tables/FS.tex
\begin{table*}[t!]
\small
\begin{center}
\begin{sc}
\begin{tabular}{llll}
\toprule
\textbf{Feature type} & \textbf{Condition} &\textbf{Critique type} \\
\midrule

missing supporting feature & 
$\Delta p_m^{\boldsymbol{x_h}+X_j} := P_M(y_h|\boldsymbol{x_h}\cup \{X_j=v_j\}) - P_M(y_h|\boldsymbol{x_h}) > \epsilon$ 
&   
Incompleteness
\\[3mm]

missing opposing feature & 
$\Delta p_m^{\boldsymbol{x_h}+X_j} := P_M(y_h|\boldsymbol{x_h}\cup \{X_j=v_j\}) - P_M(y_h|\boldsymbol{x_h}) <- \epsilon$
&   
Incompleteness
\\[3mm]

unreliable feature  &
$\Delta p_m^{\boldsymbol{x_h}-X_j} := P_M(y_h|\boldsymbol{x_h} - \{X_j=v_j\}) - P_M(y_h|\boldsymbol{x_h})>\epsilon$ 
&
Unreliability
\\[3mm]

irrelevant feature &
$|\Delta p_m^{\boldsymbol{x_h}-X_j}| := |P_M(y_h|\boldsymbol{x_h} - \{X_j=v_j\}) - P_M(y_h|\boldsymbol{x_h})|<\epsilon$ 
&
Unreliability
\\[3mm]

reliable feature &
$\Delta p_m^{\boldsymbol{x_h}-X_j} := P_M(y_h|\boldsymbol{x_h} - \{X_j=v_j\}) - P_M(y_h|\boldsymbol{x_h})<-\epsilon$ 
&
Unreliability
\\[3mm]

conflicting arguments &  
$p_m^c := P_M(y_c|\boldsymbol{x_c})>\frac{1}{C}, $ for $y_c\neq y_h, \boldsymbol{x_c}= \arg\max_{\boldsymbol{x_c^{'}}\subseteq\boldsymbol{x}} P_M(y_c|\boldsymbol{x_c^{'}})$ 
&
Conflict \\

\bottomrule
\end{tabular}
\end{sc}
\end{center}
\vspace{2pt}
\caption{
A summary of how an AI model can identify different types of issues in the decision-maker's argument using its counterfactual perspective-taking results. $\boldsymbol{x_h}$ is the decision-maker's argument, $P_M(\cdot|\cdot)$ gives the AI's confidence in a decision given an argument, and $\epsilon\in[0, 1]$ is a parameter determining whether a change in confidence is sufficiently large. 
}
\label{table:FS}
\vspace{-15pt}
\end{table*}

%% file: sections/4_experiment.tex
\section{User study}
\label{sec:experiments}

To gain a comprehensive understanding of how \aact{} impacts human-AI decision
making, especially when compared with other classical AI-based decision support systems, we designed and conducted a randomized user study on Prolific\footnote{This study was approved by the IRB of our institution.}, which serves as a case
study to answer the following research questions:

\begin{itemize}
    \item \textbf{RQ1}: How does \aact{} affect people's decision performance and their reliance on AI assistance?
    \item \textbf{RQ2}: How does \aact{} facilitate people's learning of the decision domain?
    \item \textbf{RQ3}: How does \aact{} shape people's experience, perceived critical thinking ability and perceptions of AI in decision making?
    \item \textbf{RQ4}: How do individuals with different characteristics respond to \aact{} differently?
\end{itemize}

\subsection{Task and AI Model}

In this study, we used a house sales prices prediction 
task as a case study for decision making, based on 
the Ames Housing Dataset \cite{housing}, which contained 79 features and was originally prepared for a regression problem. Similar to \cite{le2024}, we pre-processed the dataset and selected 8 features: \textit{number of bedrooms, number of central AC, number of fireplaces, overall material and finish, kitchen quality, overall condition, age when sold, living area}. 
To generate discrete class labels, we converted the house sales price into three categories, {\em Low} (less than \$100,000), {\em Medium} (\$100,000 to \$200,000), and {\em High} (more than \$200,000), resulting in an imbalanced dataset with 237, 1837, and 857 instances in each class and 2930 instances in total. In our user study, each participant was then asked to complete 20 house price prediction tasks, where in each task they were asked to review the house information and predict whether its final sales price is low, medium, or high.

We chose house sales price prediction as the decision making task in our study because it offers a realistic, interpretable domain in which laypeople may need assistance from AI but can still meaningfully reason and reflect without requiring specialized expertise. The task provides enough structure for participants to form their own argument and engage in critical reflection, while also allowing them to easily understand the information provided by the AI. Prior work in human-AI decision making has used this task~\cite{chiang2024, le2024}, demonstrating its suitability for studying how people interact with AI support in decision making.  
This task also aligns with the requirements of one of our comparison treatments, Analyzer, which necessitates a multiclass decision making problem (see details in Section~\ref{sec:treatment}).

For the AI model, we divided the Ames Housing Dataset into the training and test set based on a 80:20 split, and trained a Logistic Regression classifier on the training set as our AI model for the prediction task. On the test set, our AI model had an accuracy of 0.874.  
More details about the dataset, the pre-processing procedure, and the AI model are in Appendix~\ref{app:dataset}.

\subsection{Treatments}
\label{sec:treatment}

We included 4 treatments in our user study, comparing \textbf{\aact{}} against two established AI-based decision supports---AI as Recommender (\textbf{Recommender}) and hypothesis-driven XAI (\textbf{Analyzer})---as well as 
a \textbf{Human-only} baseline where participants make decisions with no AI assistance: 

\begin{itemize}
    \item \textbf{Human-only}: Participants make decisions on each task with no AI assistance. 
    \item \textbf{Recommender}: 
    On each task, AI assistance is provided in the form of explicit decision recommendation from the AI model, along with AI's feature importance explanations for that decision (generated by the LIME~\cite{lime} method) and confidence in that decision. 
    This is the most common AI-based decision support  used in practice~\cite{wang2021,lai2019, cai2019medical, lee2021medical, chen2025medical, bansal2021,cff, Rastogi2022anchor}, built upon explainable AI (XAI) methods (see supplementary materials Section B for example interface).

    \item \textbf{Analyzer}: On each task, 
    for {\em each} possible decision, the AI model provides evidence for and against that decision, without showing its own prediction or confidence. We again used LIME to generate supporting and opposing evidence for a decision---features with positive (or negative) importance score towards a decision are selected as the evidence for (or against) that decision (see supplementary materials Section B for example interfaces). 
    This AI-based decision support is an implementation of the hypothesis-driven XAI framework proposed by  \citeauthor{Miller2023}~\shortcite{Miller2023}, which aims to help decision-makers critically engage with AI outputs by presenting multiple hypotheses simultaneously, though it does not actively support decision-makers' reflection of their own reasoning.

    \item \textbf{\aact{}}: On each task, AI assistance is provided in the form of a guided conversation 
    between the participant and AI to critique and correct their own decision argument, as detailed in Section~\ref{sec:system} (see Figures ~\ref{fig:ui_aact1}, ~\ref{fig:ui_aact2}, ~\ref{fig:ui_aact3} and ~\ref{fig:ui_aact4} for interface examples). 
    Again, the AI model's own prediction, explanation or confidence are not shown to the participant. 
\end{itemize}

For all treatments, participants are asked to provide their decision, select a subset of features in the task as their evidence/argument for their decision, and rate their confidence in the decision in each task. For treatments with AI assistance (i.e., Recommender, Analyzer, and \aact{}), participants also have the option to update their final decision after reviewing the assistive information from AI.

\subsection{Procedure}

Upon the arrival of our study, participants first review and sign a consent form, and complete a short demographics survey including questions on their familiarity with the task background (i.e., real estate markets), familiarity with AI, need for cognition and more (see Table~\ref{table:demo} for a full list of questions in the demographic survey). 

Then, participants are directed to a task instruction page providing them a description of the house price prediction task and detailing what they are asked to do in each task. After reviewing the instruction, the participant is randomly assigned to one of the 4 treatments, and begins completing the 20 decision making tasks that are sampled from the test set of the Ames Housing dataset.

While participants in the Human-only treatment complete all 20 decision making tasks in a single session, participants in the other three treatments complete the tasks in the sequential stages: 
\begin{enumerate}
    \item \textbf{Pre-test}: Participants first complete 5 tasks without receiving any AI assistance. 
    \item \textbf{AI assistance tutorial}: Participants go through a tutorial explaining the AI assistance interface.
    For the \aact{} treatment, the tutorial explains key elements like how to update one's decision confidence in targeted self-reflection, how changes in confidence were computed (for AI-based correction suggestion and data-based triangulation), and how to interpret those changes.
    \item \textbf{Tasks with intervention}: Participants complete 10 tasks with AI assistance, and the AI assistance format is determined by the treatment assigned. 
    \item \textbf{Post-test}: Finally, participants complete 5 tasks, again without receiving any AI assistance\footnote{We made sure that in all three treatments, the AI model's accuracy in the 5 pre-test tasks, 10 intervention tasks, and 5 post-test tasks are kept the same at 80\%---closely matching the AI model's overall accuracy on the test set.}.
\end{enumerate}

Lastly, after completing all decision making tasks, participants are instructed to complete an exit survey to indicate their experience when completing the decision making tasks (e.g., cognitive load) ~\cite{nasa}, their perceived critical thinking ability in decision making~\cite{ctquestion,ctquestion2}, and their perception of AI assistance~\cite{ma2025, ma2024-3roles, chiang2024} (excluded for the Human-only treatment). These survey questions are adapted from previous literature and presented as a set of statements, where participants are asked to indicate their agreement with them on a 5-point Likert scale from 1 (strongly disagree) to 5 (strongly agree). In addition, for the three treatments with AI assistance, we also included a few open-ended questions, asking participants to comment on why they find the AI assistance they received to be helpful/not helpful. For participants in the Analyzer or \aact{} treatments, they are further asked to explicitly indicate whether they prefer the AI assistance they've received or an alternative option where the AI model directly provides decision recommendation (i.e., the ``Recommender'') and why. We also include an attention check question in the exit survey, asking participants to select a pre-specified option. The full list of exit survey questions can be found in Table~\ref{table:survey}.

The study was conducted on Prolific and restricted to U.S.-based workers. Each participant could take the study only once and received \$3.80 upon completion. To incentivize participants' best decision performance, we provided participants who achieved an 80+\% accuracy in the 20 decision making tasks a \$1 bonus. In the end, the study's median completion time was 25 minutes, yielding an effective hourly pay of \$8.94 (\$11.29 for participants who received the bonus).

\subsection{Measurements}
Corresponding to our research questions, we defined metrics that measure four key dimensions: \textit{decision performance}, \textit{AI reliance}, \textit{learning}, and \textit{subjective perceptions}.

\begin{itemize}
    \item \textbf{Decision performance}: 
    We used the \textit{accuracy} of participants' final decision in tasks 6--15 (where AI assistance intervention may be available) as our main metric to measure decision performance. 
    Since our dataset is imbalanced, we also used \textit{balanced accuracy} on these tasks, defined as the average of recall for each class, to complement the results. 
    
    \item \textbf{Reliance}: Following previous literature on human-AI decision making~\cite{ma2024-3roles,DKE,bansal2021, wang2021}, we used 
    \textit{agreement fraction} and \textit{switch fraction} as general measurements on how often humans follow AI advice. 
    We also used \textit{over-reliance} and \textit{under-reliance ratios} to measure the appropriateness of humans' reliance towards on AI;
    over-reliance is the frequency at which participants follow AI's incorrect decision while under-reliance is the frequency that participants fail to follow AI's correct decision.
    Reliance metrics are computed for tasks 6--15\footnote{We computed these metrics based on the withheld AI predictions for treatments where AI predictions are not directly shown (i.e., Analyzer and \aact{}).}
    and metric details are provided in Table~\ref{table:metric}.

    \item \textbf{Learning}: 
    We defined two metrics to capture possible domain knowledge learning that may occur at different stages: 
    \begin{itemize}
        \item \textit{Learning during intervention}: We measured the change in the participant's  accuracy between the their \textit{initial} decision in the intervention stage (tasks 6--15) and the pre-test stage (tasks 1--5). This reflects how participants' ability to solve the decision task independently evolves while engaging with AI assistance.
        \item \textit{Learning after intervention}: We measured the change in the participant's accuracy between the post-test (tasks 16--20) and the pre-test (tasks 1--5) stage. 
        This shows how AI assistance can prepare decision-makers to solve tasks independently without future support. 
    \end{itemize}
    
    Following previous literature~\cite{normalizedchange,Gajos2022}, we measured normalized change to reward improvements from high baselines. Metric details are provided in Table~\ref{table:metric}.
     
    \item \textbf{Subjective Perceptions}: Our subjective perception measurements were directly taken from participants' responses in the exit survey (i.e., reported on a 5-point Likert scale), and can be divided into three main categories:
    \begin{itemize}
    \item {\em User experience}: primarily concerning participants' confidence and cognitive load in decision making as measured by the NASA TLX scale~\cite{nasa}, including sub-components mental demand, effort, and performance/efficacy.
    \item {\em Perceived critical thinking abilities}: participants' self-reported ability to perform critical thinking in decision making, including their ability to conduct comprehensive evidence evaluation, consider multiple perspectives, actively seek counter-evidence, and explain their own decisions.
    \item {\em AI perceptions}: Participants' self-reported perceptions of the AI assistance that they received in their treatment, regarding AI's helpfulness, trustworthiness, understandability, and ability to provoke deep thinking and reflection. Participants also rated on their decision autonomy when receiving AI assistance, their satisfaction with the AI assistance, and their willingness to use the AI assistance in the future.
    \end{itemize}

\end{itemize}

\subsection{Analysis Methods}

To answer \textbf{RQ1--RQ3}, for each of the corresponding measures, we start by performing an one-way ANOVA test across all treatments followed by post-hoc pairwise comparisons using Tukey's HSD test. 
For pairs of treatments with significant differences ($p<0.05$), we report the effect size with Cohen's $d$.
As our main interest is in understanding the effects of \aact{} on various measurements, in the paper, we mainly report pairwise comaprisons between the \aact{} treatment and other treatments.
To answer \textbf{RQ4}, 
we divided participants into subgroups based on their demographics characteristics (e.g., their task familiarity and AI familiarity), and repeated the analyses above within each subgroup. 

Moreover, to gain qualitative insights into why decision-makers may like or dislike \aact{}, we conducted an exploratory analysis
of the open-ended responses from participants in the \aact{} treatment, when they were asked to compare \aact{} with an alternative format of AI assistance where the AI model directly provides decision recommendations. The first author of the paper performed open coding of these responses, which were then clustered into higher-level themes following inductive thematic analysis methods~\cite{hsieh2005three}. We then reported the common themes identified along with some example responses.

%% file: sections/5_results.tex
\section{Results}
\label{sec:results}

In total, 405 Prolific workers participated in our study. 
2 participants did not complete all 20 decision tasks and 1 failed the attention check question, leaving us the valid data from 402 participants (Human-only: 104, Recommender: 97, Analyzer: 97, \aact{}: 104). We include the demographic details of our participants in Appendix~\ref{app:demo}. 
We then analyze this data to answer our research questions. Since no significant results are found for RQ2 (effects on learning), we leave it to Appendix~\ref{app:RQ2} and focus on the other three RQs here.

\subsection{RQ1: Effects on Decision Performance and AI Reliance}
\label{sec:RQ1.1}

\subsubsection{Decision Performance}

\begin{figure*}[t!]
    \centering
    \begin{subfigure}[b]{0.35\textwidth}
        \includegraphics[width=\textwidth]{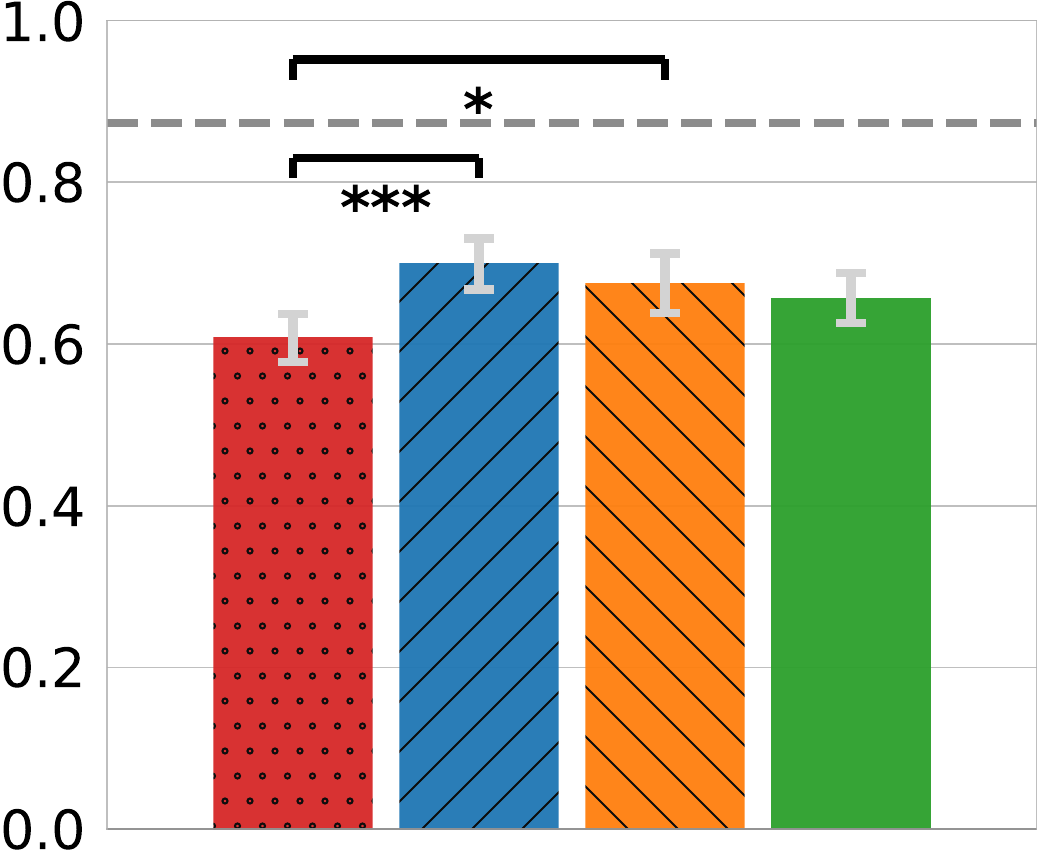}
        \caption{Accuracy}
        \label{fig:performanceA}
    \end{subfigure}
    \hspace{5mm}
    \begin{subfigure}[b]{0.35\textwidth}
        \includegraphics[width=\textwidth]{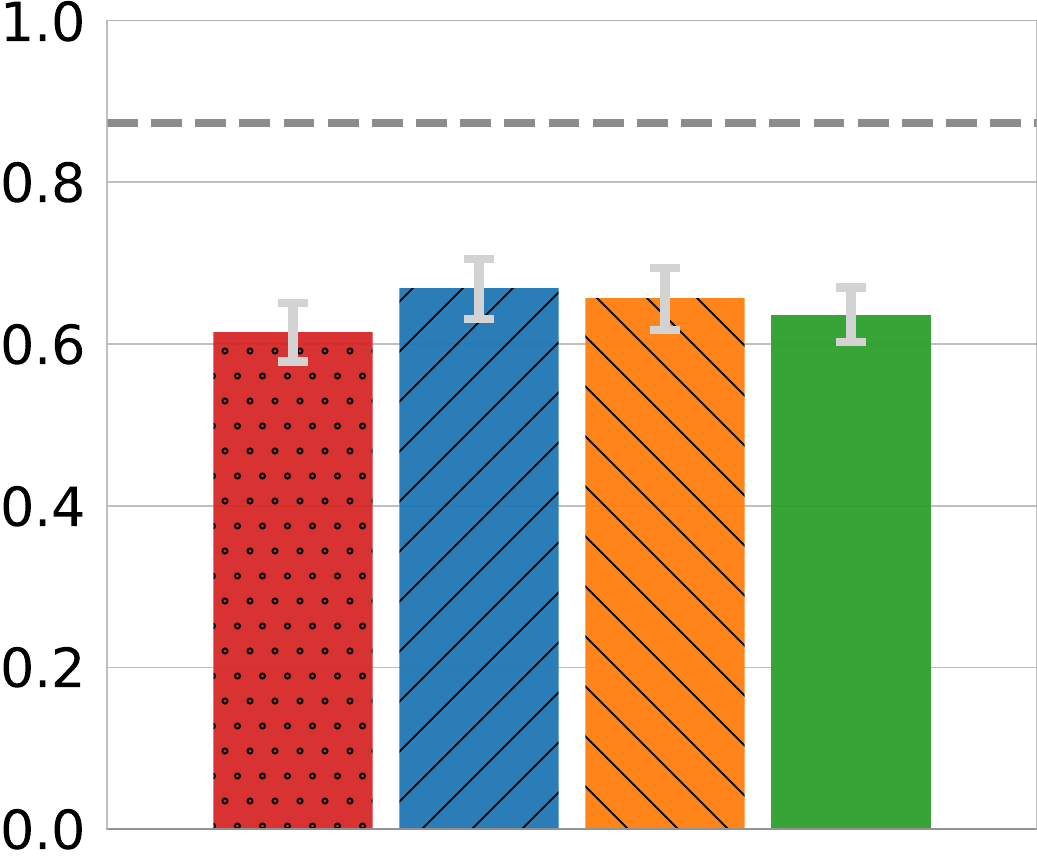}
        \caption{Balanced Accuracy}
        \label{fig:performanceB}
    \end{subfigure}
    \begin{subfigure}[b]{0.18\textwidth}
        \includegraphics[width=\textwidth]{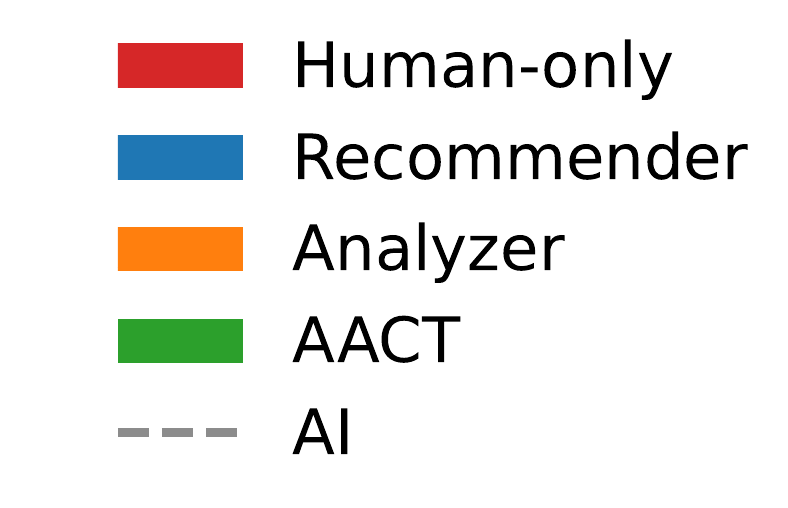}
        \vspace{12mm}
    \end{subfigure}
    \caption{Comparison on participants' decision performance measured by their (a) accuracy and (b) balanced accuracy. Error bars represent the 95\% confidence intervals of the mean values. $\textsuperscript{*}$, $\textsuperscript{**}$, and $\textsuperscript{***}$ denote statistical significance levels of $0.05$, $0.01$, and $0.001$ respectively.}
    \label{fig:performance}
\end{figure*}

Figure ~\ref{fig:performance} compares participants' decision accuracy and balanced accuracy in tasks 6--15 across all four treatments. 
The one-way ANOVA test suggests there is a significant difference across treatments in participants' decision accuracy  ($F(3,398)=5.704, p=0.001, \eta^2=0.041$). Pairwise comparison results suggest the difference mainly came from the improvement brought up by the Recommender and Analyzer treatments over the Human-only treatment,   
and there was no significant differences between \aact{} and other treatments.
Interestingly, when focusing on participants' balanced accuracy in tasks 6--15, we find no significant difference across treatments ($F(3,398)=1.643, p=0.179, \eta^2=0.012$). This implies that compared to the Human-only treatment, various forms of AI assistance, including \aact{}, may only improve participants' overall decision performance but not their performance on those tasks with minority labels.

\begin{figure*}[t!] 
    \centering
    \begin{subfigure}[b]{0.79\textwidth}
        \includegraphics[width=\textwidth]{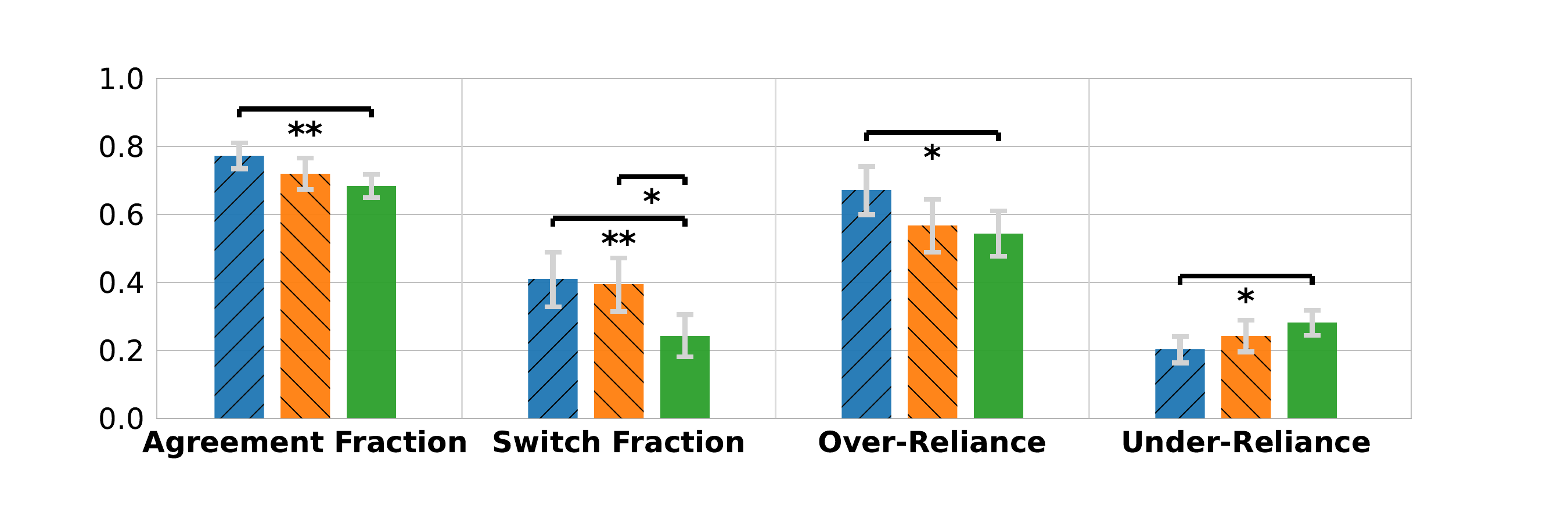}  
    \end{subfigure}
    \hspace{-40pt}
    \begin{subfigure}[b]{0.2\textwidth}
        \includegraphics[width=\textwidth]{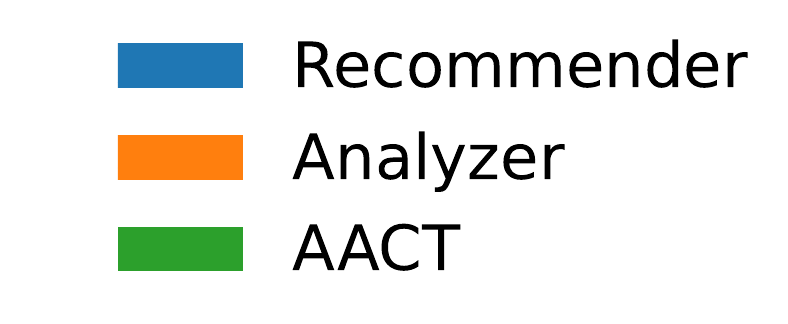}
        \vspace{10mm}
    \end{subfigure}
    \vspace{-10pt}
        \caption{Comparisons on participants' reliance (agreement and switch fraction) and appropriateness of reliance (over-reliance ratio and under-reliance ratio) on AI. Error bars represent the 95\% confidence intervals of the mean values. $\textsuperscript{*}$, $\textsuperscript{**}$, and $\textsuperscript{***}$ denote statistical significance levels of $0.05$, $0.01$, and $0.001$ respectively.}
        \label{fig:reliance}    
    
\end{figure*}

\subsubsection{Reliance on AI}
\label{sec:RQ1.2}

Figure ~\ref{fig:reliance} compares participants' reliance on AI in tasks 6--15 across the three treatments with AI assistance.  
We first focus on agreement and switch fractions, the two metrics capturing participants' general tendency to rely on the AI model's prediction. 
The one-way ANOVA tests yielded significant differences across treatments for both metrics (agreement fraction: $F(2,295)=4.974, p=0.008, \eta^2=0.033$;  switch fraction: $F(2,295)=6.222, p=0.002, \eta^2=0.04$).
The pairwise comparisons further suggested that participants agreed with AI's predictions significantly less in the \aact{} treatment ($M=0.684, SD=0.177$) than those in the Recommender treatment ($M=0.772, SD=0.188; p=0.005$, Cohen's $d=-0.485$). In cases where participants' initial decision differed from the AI model's prediction, participants also switched to AI's prediction significantly less in the \aact{} treatment ($M=0.243, SD=0.319$) compared to those in the Recommender ($M=0.409, SD=0.398; p=0.005$, Cohen's $d=-0.461$) or Analyzer ($M=0.393, SD=0.389; p=0.012$, Cohen's $d=-0.423$) treatments.

We then move on to examine participants' appropriateness in their reliance on AI through over-reliance 
and under-reliance ratios (lower is better).
One-way ANOVA tests again revealed significant differences across treatments for both metrics (over-reliance ratio: $F(2,295)=3.462, p=0.033, \eta^2=0.023$; under-reliance ratio $F(2,295)=3.681, p=0.026, \eta^2=0.024$).
Moreover, we find that compared to participants in the Recommender treatment ($M=0.67, SD=0.353$), those in the \aact{} treatment ($M=0.543, SD=0.342$) relied on AI significantly less when AI's prediction was wrong, leading to a significantly lower level of over-reliance on AI ($p=0.035$, Cohen's $d=-0.365$). In contrast, participants of the \aact{} treatment ($M=0.281, SD=0.191$) relied on AI significantly less than those in the Recommender treatment ($M=0.202, SD=0.195$) when AI's prediction was correct, leading to a significantly higher level of under-reliance on AI ($p=0.019$, Cohen's $d=0.409$). In other words, our results indicate that \aact{} is highly effective in reducing decision-makers' over-reliance on AI, but it may also come at a cost of increasing their under-reliance on AI.

\begin{figure*}[t] 
\begin{flushleft}
    \begin{subfigure}[b]{0.84\textwidth}
        \includegraphics[width=\textwidth]{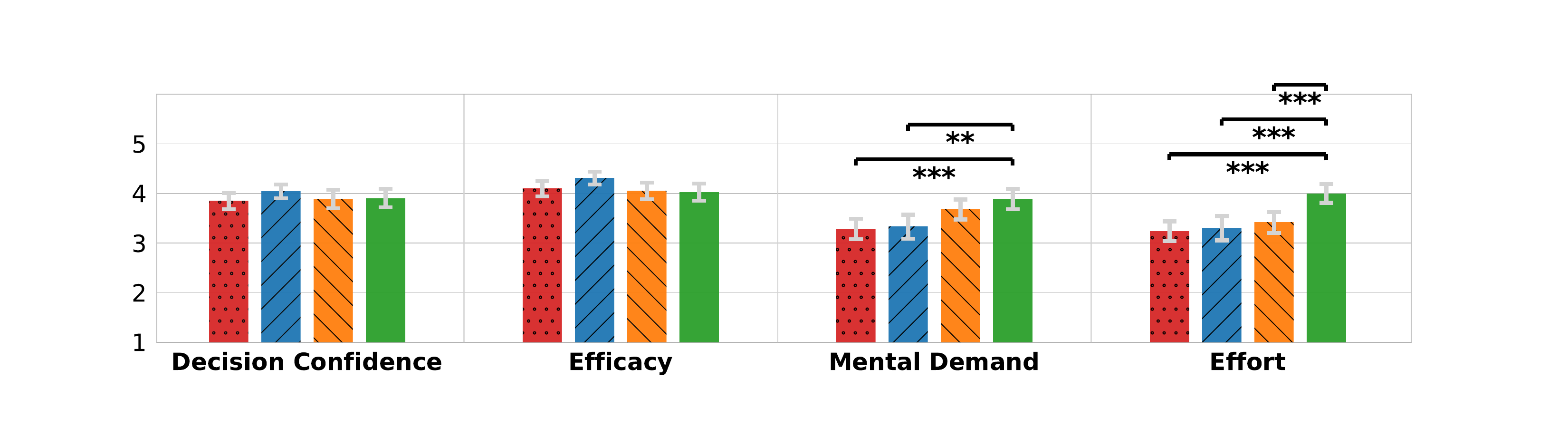}  
        \vspace{-7mm}
        \caption{User experience}
        \label{fig:surveyA}
    \end{subfigure}
    \\[-7mm]
    \begin{subfigure}[b]{0.84\textwidth}
        \includegraphics[width=\textwidth]{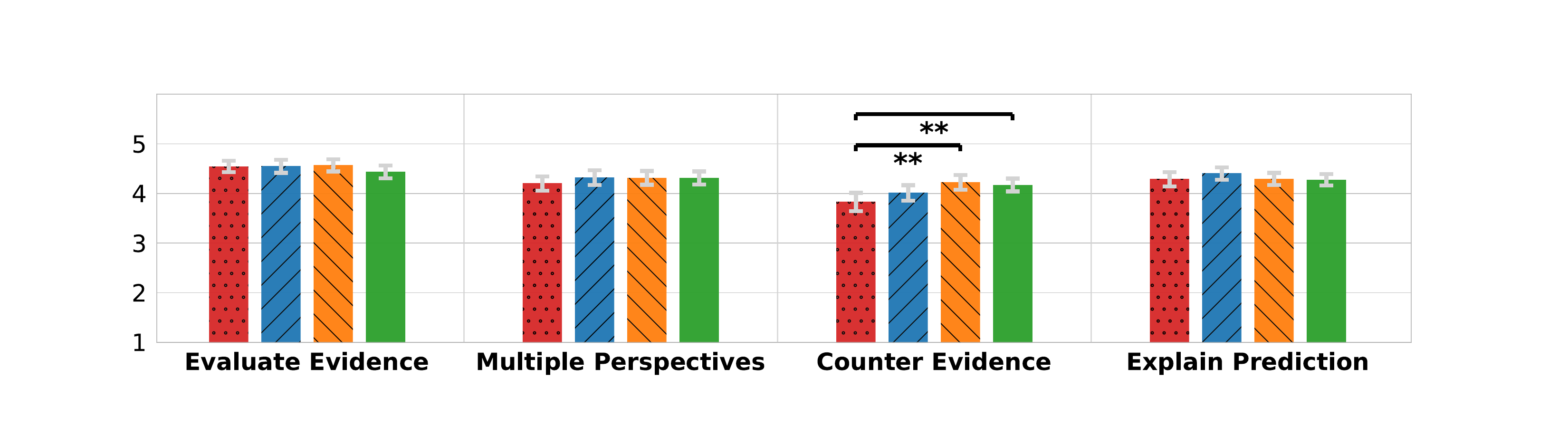}
        \vspace{-7mm}
        \caption{Perceived critical thinking abilities}
        \label{fig:surveyB}
    \end{subfigure}
    \hspace{-12mm}
    \begin{subfigure}[b]{0.15\textwidth}
        \includegraphics[width=\textwidth]{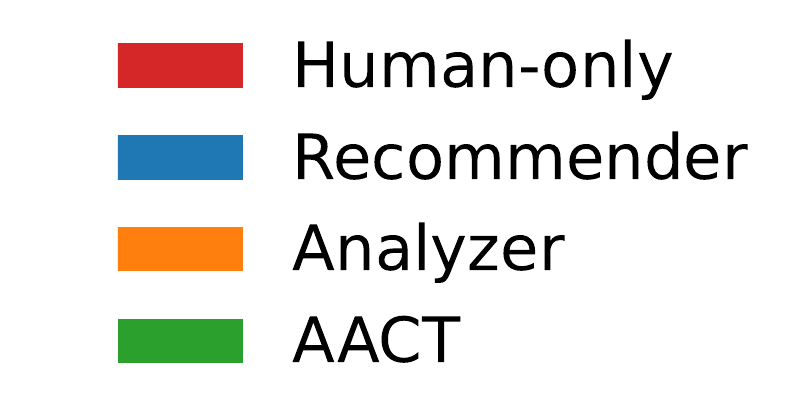}
        \vspace{33mm}
    \end{subfigure}
    \caption{Comparisons on users' subjective perceptions  across treatments. Error bars represent 95\% confidence intervals of the mean values.  $\textsuperscript{**}$and $\textsuperscript{***}$ denote statistical significance levels of $0.01$ and $0.001$ respectively.}
    \label{fig:survey}
\end{flushleft}
\end{figure*}

\subsection{RQ3: Effects on User Experience, Perceived Critical Thinking Abilities, and Perceptions of AI}
\label{sec:rq3}

We now examine the impacts of \aact{} on participants' subjective experience during decision making.
Figures~\ref{fig:surveyA} and~\ref{fig:surveyB} compares various user experience metrics and the perceptions of critical thinking abilities, respectively, for participants in different treatments. 
It is shown that participants in the \aact{} treatment report significantly higher mental demand ($M=3.887, SD=1.054$) compared to participants in Human-only ($M=3.286, SD=1.072$; $p<0.001$, Cohen's $d=0.566$) and  Recommender ($M=3.33, SD=1.205$; $p=0.002$, Cohen's $d=0.493$) treatments.
Similarly, \aact{} participants also report  higher effort to accomplish their level of performance 
compared to participants in 
all three other treatments (Human-only: $p<0.001$, Cohen's $d=0.755$; Recommender: $p<0.001$, Cohen's $d=0.638$; Analyzer: $p=0.001$, Cohen's $d=0.571$).
Regarding participants' perceived critical-thinking abilities during decision making, while \aact{} participants reported similar levels of abilities in evaluating evidence comprehensively, examining multiple perspectives, and explaining their own decisions as other participants, they also reported a higher likelihood to seek evidence that counter their own knowledge compared to participants in the Human-only treatment ($p=0.01, d=0.417$). This may because the \aact{} system 
actively suggests counter-evidence to decision-makers both when it identifies incompleteness and conflict issues in their arguments.

Finally, for participants' perceptions of AI assistance, we observe no significant differences between treatments and leave the results to Appendix~\ref{app:RQ3}.

\subsection{RQ4: Individual Differences of the Effects of \aact{}}
\label{sec:demo}

To understand how individuals with different characteristics may respond to \aact{} differently (RQ4), we move on to repeat our analyses for RQ1--RQ3 within different subgroups of participants defined by various demographic dimensions.  
We collected 5 types of demographics data through a pre-task survey: participants' task familiarity, AI familiarity, Need for Cognition (NFC) \cite{nfc}, Cognitive Reflection Test (CRT) scores \cite{crt}, and education levels; we did not find strong correlation between any pair of these demographics.
In this subsection, we highlight findings that 
indicate which subgroup may experience the greatest benefits or costs from \aact{}, with additional results in Appendix~\ref{app:RQ4}.

\subsubsection{AI familiarity, task familiarity, and education impact whether individuals can reduce over-reliance on AI through \aact{}}

\begin{figure*}[t]
    \begin{subfigure}[b]{0.26\textwidth}
        \includegraphics[width=\textwidth]{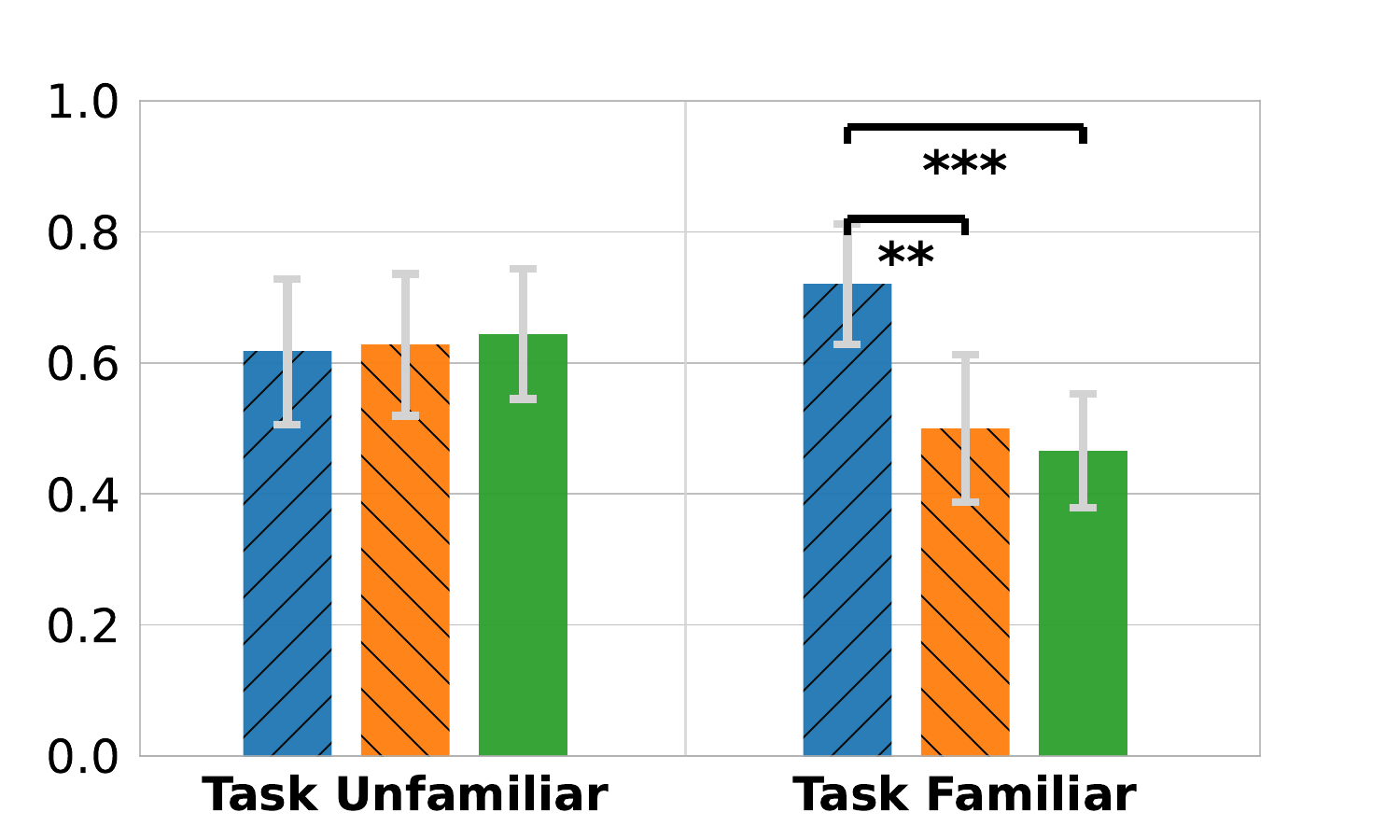}
        \caption{Task Familiarity}
        \label{fig:overrelianceA}
    \end{subfigure}
    \begin{subfigure}[b]{0.34\textwidth}
        \includegraphics[width=\textwidth]{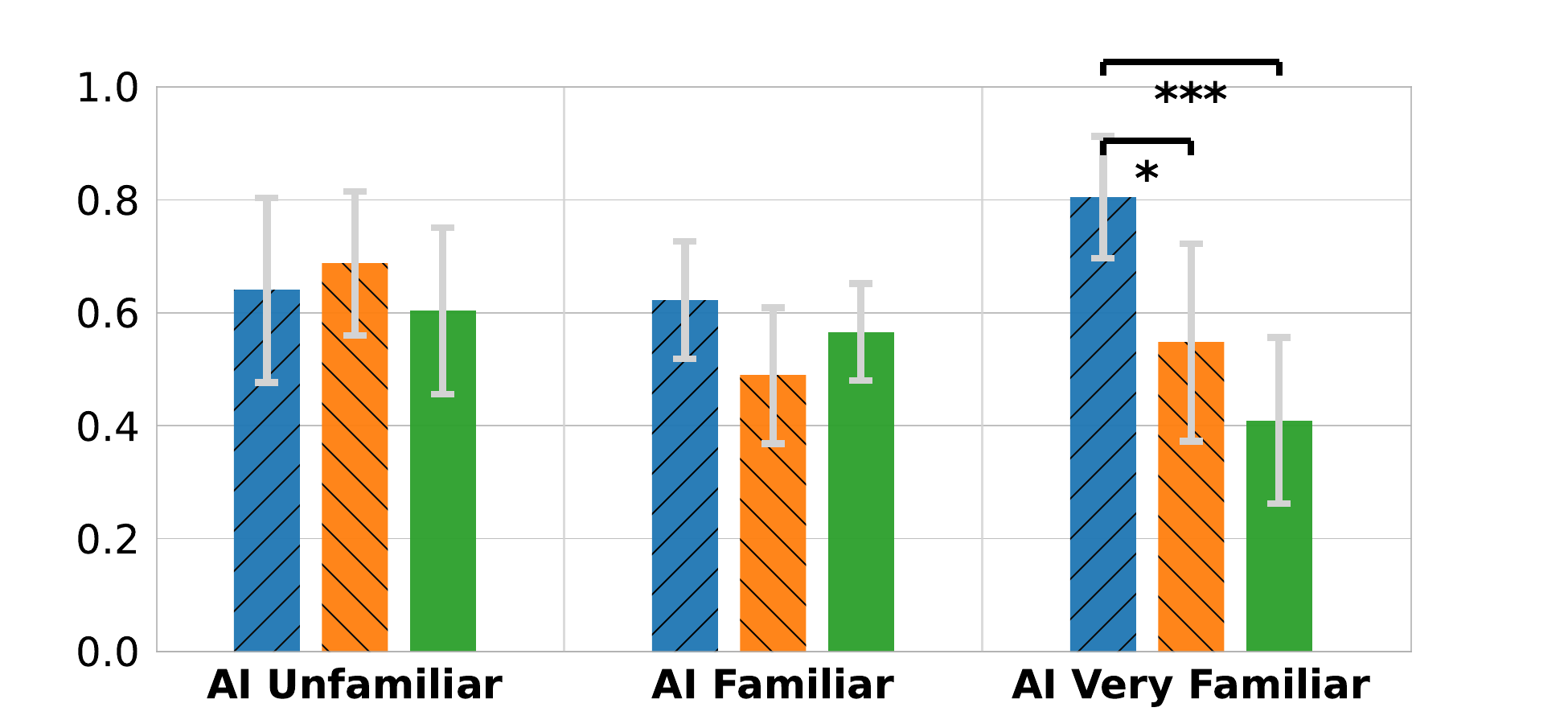}
        \caption{AI Familiarity}
        \label{fig:overrelianceB}
    \end{subfigure}
    \begin{subfigure}[b]{0.26\textwidth}
        \includegraphics[width=\textwidth]{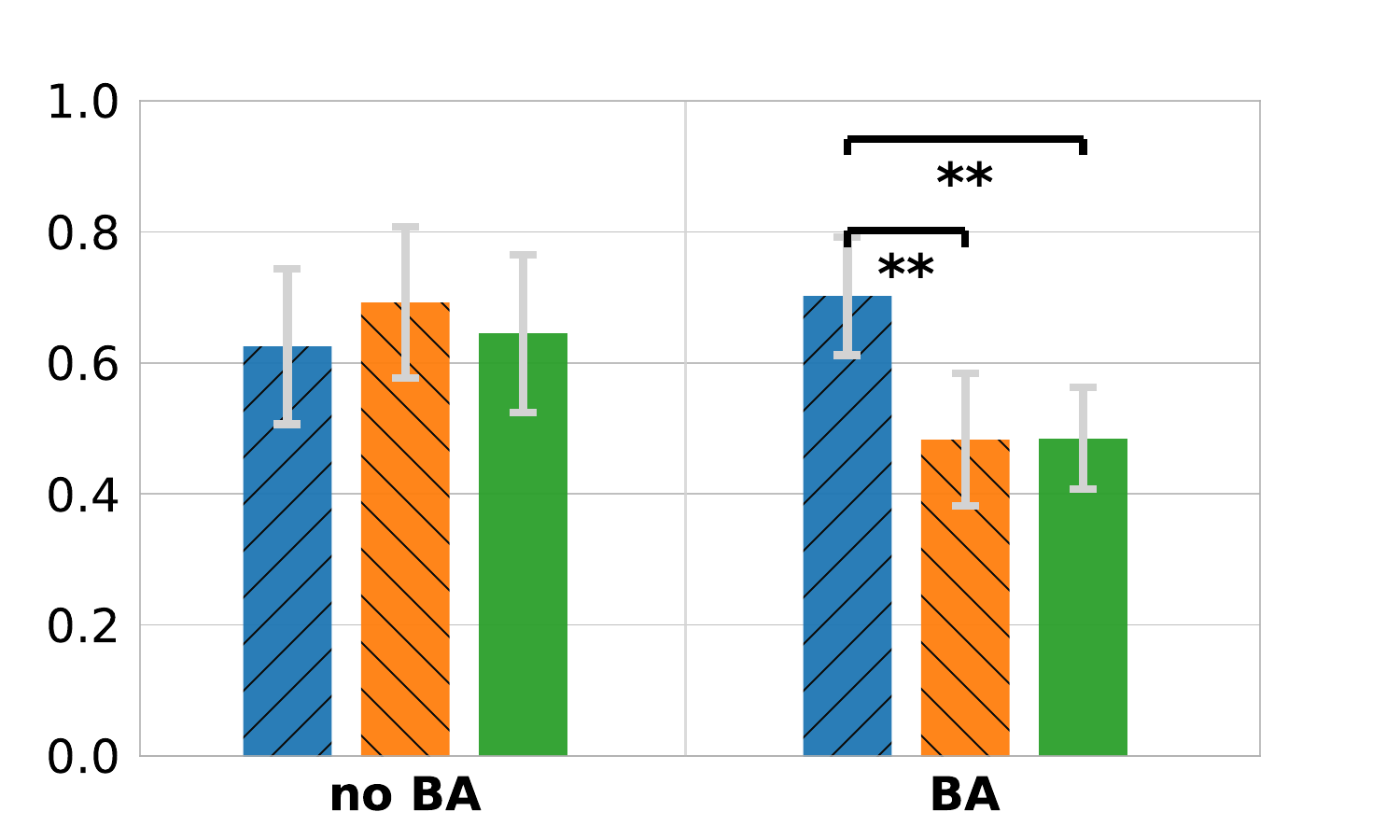}
        \caption{Education}
        \label{fig:overrelianceD}
    \end{subfigure}
    \begin{subfigure}[b]{0.12\textwidth}
        \includegraphics[width=\textwidth]{figs/figs_main/legends/legend.pdf}
        \vspace{8mm}
    \end{subfigure}
    \caption{Over-reliance ratio for subgroups devided by (a) task familiarity, (b) AI familiarity and (c) education. Error bar represent 95\% confidence intervals of the mean values. $\textsuperscript{*}$, $\textsuperscript{**}$ and $\textsuperscript{***}$ denote statistical significance levels of $0.05$, $0.01$ and $0.001$ respectively.}
    \label{fig:overreliance}
\end{figure*}

We have previously found that \aact{} can be highly effective in helping reducing people's over-reliance on AI. Our individual differences analysis, however, reveals that such positive effect may only occur within certain subgroups. For example, Figure~\ref{fig:overreliance} presents the over-reliance ratios for subgroups of participants with different task familiarity, AI familiarity, and education level. We find that among participants who reported lower levels of familiarity with the task background of real estate markets (i.e., the ``task unfamiliar'' subgroup who selected options 1--2 ``not at all/slightly familiar''; $n=195$), there is no difference in their over-reliance on AI when they interact with different types of AI supports. However, for participants who reported higher levels of task familiarity (i.e., the ``task familiar'' subgroup who selected options 3--5 ``somewhat/moderately/extremely familiar''; $n=207$), we observe significant differences across treatments on participants' over-reliance on AI ($F(2,152)=8.292, p<0.001, \eta^2=0.098$), with those in the \aact{} treatment over-relied on AI significantly less than those in the Recommender treatment ($p=0.001$, Cohen's $d=-0.774$). This implies that only individuals who are more familiar with the task background may benefit from \aact{} to reduce their over-reliance on AI. Indeed, further t-test between participants in the \aact{} treatment who had different levels of task familiarity also confirmed that the task familiar subgroup achieved significantly lower over-reliance on AI than the task unfamiliar subgroup ($p=0.008$, Cohen's $d=-0.537$).

Similarly, we also find that only among participants who reported to be very familiar with AI (selected option 5, $n=93$) we detected significant differences in over-reliance ration across treatments ($F(2,63)=8.607, p<0.001, \eta^2=0.215$), with \aact{} users over-relied on AI significantly less than Recommender users ($p<0.001$, Cohen's $d=-1.35$). Finally, as shown in Figure~\ref{fig:overrelianceD}, \aact{} only enabled participants with bachelor degrees or above  ($n=240$) to significantly reduce their over-reliance on AI ($F(2,178)=7.711, p=0.001, \eta^2=0.08$; \aact{} vs. Recommender: $p=0.002$, Cohen's $d=-0.664$) but not those with lower education levels.

\begin{figure*}[t!]
    \begin{subfigure}[b]{0.43\textwidth}
        \includegraphics[width=\textwidth]{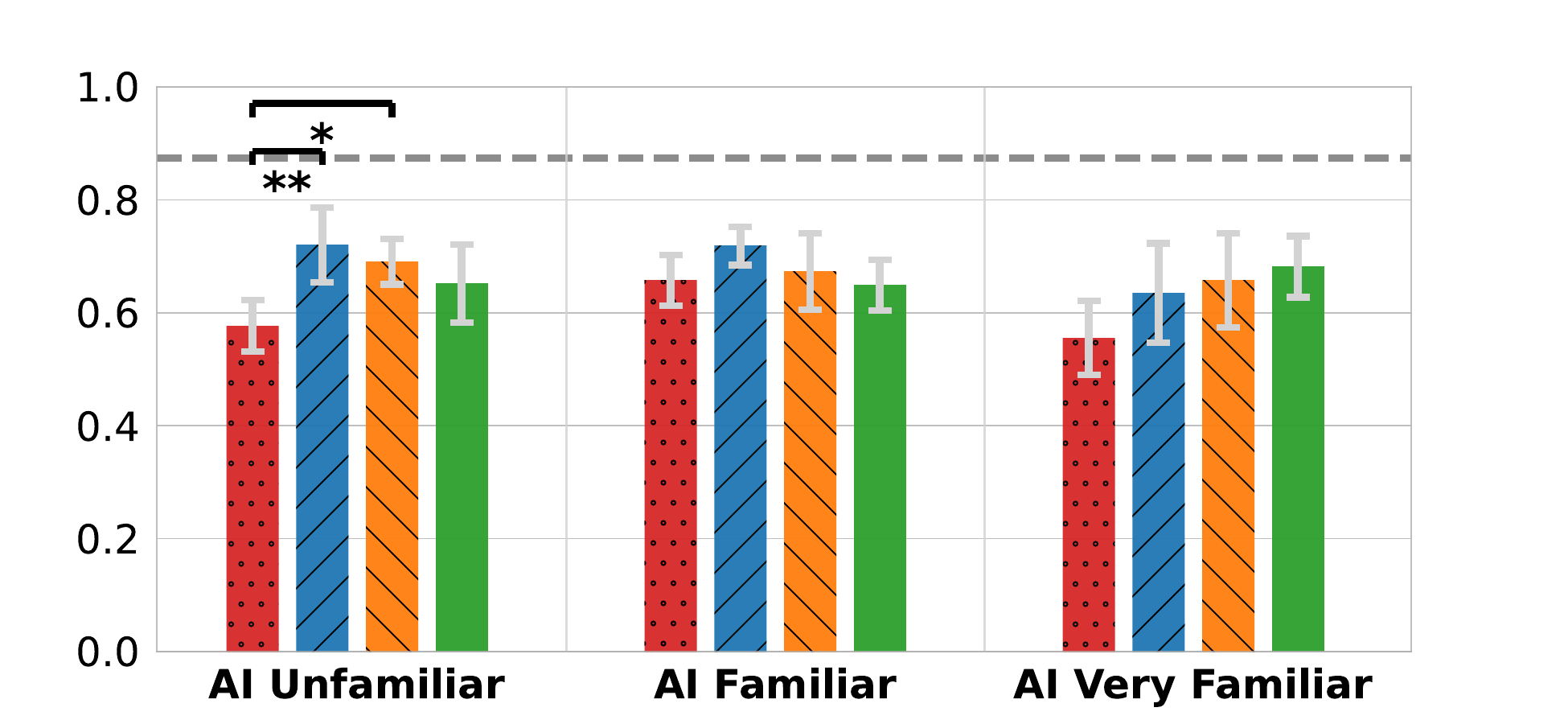}
        \caption{Accuracy}
        \label{fig:AI_famA}
    \end{subfigure}
    \begin{subfigure}[b]{0.43\textwidth}
        \includegraphics[width=\textwidth]{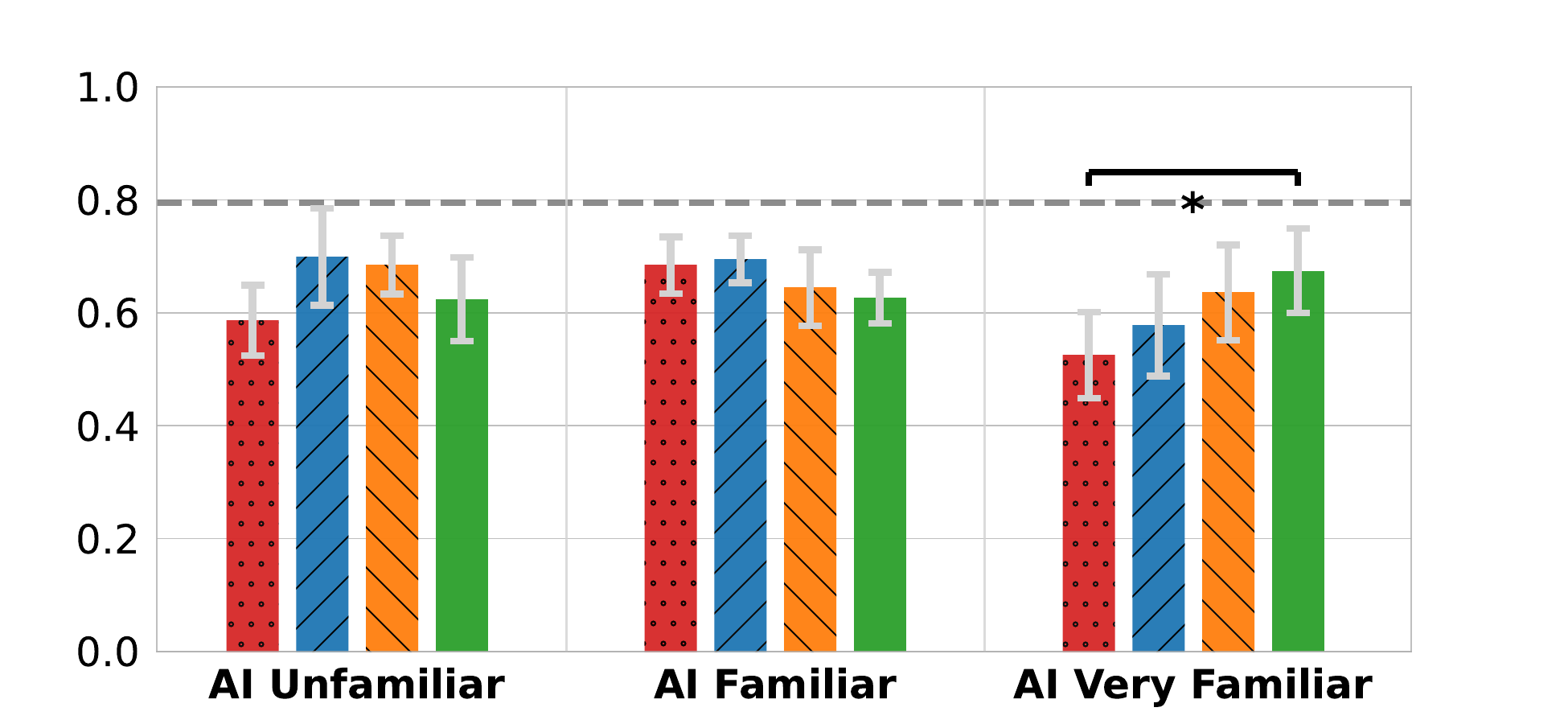}
        \caption{Balanced Accuracy}
        \label{fig:AI_famB}
    \end{subfigure}
    \begin{subfigure}[b]{0.13\textwidth}
        \includegraphics[width=\textwidth]{figs/figs_main/legends/legend_h_AI.pdf}
        \vspace{10mm}
    \end{subfigure}
    \caption{Decision performance (a: accuracy; b: balanced accuracy) for subgroups with different AI familiarity. Error bar represent 95\% confidence intervals of the mean values. $\textsuperscript{*}$ and $\textsuperscript{**}$ denote statistical significance levels of $0.05$ and $0.01$  respectively.}\label{fig:ai:accuracy}
\end{figure*}

\subsubsection{\aact{} may help individuals with very high AI familiarity improve decision performance}

Figure~\ref{fig:ai:accuracy} compares the decision performance across the four treatments for participants who reported unfamiliar (selected options 1--3, $n=116$), familiar (selected option 4, $n=193$) or very familiar with AI (selected option 5, $n=93$). While the trends in the AI unfamiliar and AI familiar subgroups largely resemble those in the general population (i.e., Figure~\ref{fig:performance}), the trend among participants who are very familiar with AI is different---for these participants, having them interact with \aact{} seem to result in the highest level of decision performance. Indeed, for the AI very familiar subgroup, we find a marginally significant difference across treatments on decision accuracy $F(3,89)=2.557, p=0.06, \eta^2=0.079$, and a significant difference across treatments on balanced accuracy $F(3,89)=2.88, p=0.04, \eta^2=0.088$. Pairwise comparison also indicated that when participants are very familiar with AI, interacting with \aact{} helps them achieve a significantly higher level of balanced accuracy than those who complete decisions alone ($p=0.037$, Cohen's $d=0.819$). 
Interestingly, we note that these differences are mainly caused by that participants who reported to be very familiar with AI tend to have worse decision performance than other subgroups when receiving no AI assistance (one-way ANOVA for accuracy: $F(2,101)=5.029, p=0.008, \eta^2=0.091$; for balanced accuracy: $F(2,101)=7.554, p=0.001, \eta^2=0.13$) or direct decision recommendations from AI (e.g., one-way ANOVA on balanced accuracy: $F(2,94)=3.838, p=0.025, \eta^2=0.076$)---the latter may result directly from their over-reliance on AI (see Figure~\ref{fig:overrelianceB}). In this sense, \aact{} may be particularly helpful for decision-makers with very high familiarity with AI, as it can help them reflect more critically of AI information and improve their decision performance. This is also consistent with the participants' subjective perceptions---for example, among participants who are very familiar with AI, the \aact{} users perceived AI as significantly more reflection encouraging than the Recommender users ($p=0.009$, Cohen's $d=0.819$).

\subsection{Exploratory Analysis: Why did Users Like or Dislike \aact{}?}

Finally, by analyzing the open-ended responses from participants in the \aact{} treatment on their comparison of the 
\aact{} system they used in the study with direct decision recommendations from AI (i.e., Recommender), we revealed a few interesting reasons underlying people's preferences. 

102 of the 104 participants in the \aact{} treatment provided valid responses. Overall, 
75 preferred Recommender, 22 preferred \aact{}, and 5 had no preferences. 
Primary reasons for participants to prefer \aact{} include increased decision autonomy (e.g., \textit{``Now I feel like I have agency over my decision while taking advantage of AI ideas''}), capability to think independently before seeing AI's information (e.g., \textit{``This method motivates me to think for myself first, then reflect on and enhance my own reasoning using the AI's suggestions''}), and promotion of deeper thinking (e.g., \textit{``it helped me use my brain and think about why I was making decisions''}).
Similarly, people may not prefer Recommender as they worry being biased by AI (e.g., \textit{``If the AI predictions are shown directly, that could bias me too much toward the AI choice''}).

On the other hand, we also note that participants' main reasons for disliking \aact{} included information overload and confusion (e.g., \textit{``It gave too much info and that confused me''}) and heightened self-doubt 
(\textit{``I think [\aact{}] made me doubt my answers more''}).
Meanwhile, participants who preferred Recommender mainly wanted to get AI predictions to compare against their own and preferred its simplicity and straightforwardness.

%% file: sections/6_discussion.tex
\section{Discussions}
\label{sec:discussion}

We introduced the \textit{AI-Assisted Critical Thinking} framework as an AI-based decision support system that helps decision-makers critique and correct their own decision.
Our findings indicate that it can be helpful for reducing over-reliance towards AI and avoiding AI errors, especially for certain subgroups, but it also comes with costs such as increased cognitive load. 
Below, we discuss implications of our findings and future design considerations. We also acknowledge the generalizability and limitations of our study and provide directions for future work.

\input{tables/benefit_costs}

\subsection{Benefits, Costs and Use Cases  of the Current \aact{} System}
\label{sec:current}
Table~\ref{tab:aact_benefits_costs} summarizes the key benefits and costs of the current instantiation of the \aact{} system, as observed in our user study, which reveals a clear trade-off: \aact{} decreases over-reliance on AI and strengthens decision autonomy and metacognitive rigor, but at the cost of increased cognitive effort and a greater risk of under-reliance. In our study, participants using \aact{} showed markedly reduced over-reliance on AI. This effect arises directly from how \aact{} structures reflection: the system never reveals the AI model's own prediction; the targeted self-reflection required before viewing the AI's feature-level correction suggestions preserves a degree of independence in decision-makers' evaluation of the internal coherence of their own argument; and the data-based triangulation offers an additional reference point for assessing the trustworthiness of AI suggestions. At the same time, this deeper engagement 
naturally increases cognitive load and slows the decision process. 
The heightened critical stance also makes decision-makers more cautious toward information from AI overall, leading to an increase of under-reliance even when the AI is correct. In this sense, \aact{} trades efficiency for autonomy and robustness---helping decision-makers think harder and rely less automatically on AI, but at the cost of greater mental effort and a generally more critical posture toward AI assistance.

These trade-offs provide implications on where \aact{} may be most and least beneficial. \aact{} could be well-suited for domains where reflective reasoning and decision autonomy is central, such as medical diagnosis, financial auditing, or other contexts where justifying one's rationale is as important as the final decision. 
Given \aact{}'s strong effects in reducing over-reliance on AI, it may also be especially valuable when the AI model's independent performance is limited, or in high-stakes settings where blindly following an incorrect AI recommendation would be costly. In contrast, for time-sensitive or low-stakes tasks supported by highly reliable AI systems, directly providing AI recommendations and explanations or using light-touch reflection-oriented approaches centered around AI outputs (e.g., hypothesis-driven XAI) may offer a better cost-benefit balance.

In addition, our subgroup analysis sheds light on which decision-makers may benefit most from \aact{}. We found that participants with greater self-reported task knowledge, higher self-perceived familiarity with AI, and higher levels of education exhibited lower over-reliance when using \aact{} compared to other forms of AI assistance.
One explanation is that \aact{} presents feedback at a granular, feature-level resolution, and decision-makers with stronger domain knowledge or higher education are both less prone to information overload and better able to interpret the probabilistic and counterfactual information the system provides. 
Meanwhile, participants who reported high familiarity with AI tended to over-rely on AI in the baseline conditions, likely due to stronger default trust in AI systems. \aact{} mitigated this tendency by redirecting their attention to the structure of their own reasoning rather than the AI’s prediction.
Taken together, these findings  suggest that \aact{} may be more suitable for decision-makers who can comfortably engage in metacognitive reasoning, such as domain experts, as well as for those who are susceptible to over-reliance on AI, such as users with high dispositional trust in AI.

Finally, although our study did not directly compare \aact{} with some recent reflection-provoking approaches that leverage LLMs for evaluating or critiquing human reasoning~\cite{khadar2025wisdom,reicherts2025ai}, we note that a key distinction between them lies in how reflections are generated. 
\aact{} evaluates decision-makers' argument using a \textit{domain-specific model}, enabling it to flag the most task-relevant flaws in the decision-maker's argument based on domain-relevant evidence patterns. This ensures that the reflection it elicits are meaningful for the specific task. In contrast, LLM-generated prompts may surface counterexamples or critiques that are linguistically plausible but not necessarily aligned with the domain's underlying data distribution, potentially leading to reflections that feel arbitrary or misdirected. As such, \aact{} can be better suited for structured decision tasks, especially for domains where LLM's world knowledge is still lacking, while LLM-based interrogation can be more appropriate for open-ended reasoning or general critical-thinking support.

\input{tables/design}

\subsection{Design Implications and System Scalability}
\label{sec:design}
\subsubsection{Improving the Scalability of \aact{} via Adaptive Designs} Section~\ref{sec:dc} outlines the design space underlying the \aact{} framework, and our user study illustrates how specific instantiation of the design choices shape both its benefits and costs. As discussed in  Section~\ref{sec:current}, the current implementation of \aact{} offers strong gains in autonomy and reduced over-reliance but also imposes substantial cognitive demand. These findings point to concrete opportunities for scaling and simplifying \aact{} in real-world deployments. Table~\ref{table:design} summarizes several design knobs that can be adjusted to lower the cognitive burden of \aact{} and increase its practical applicability. A key strategy is to support the ``quick-test'' component from the R/M model, which shortens the duration of metacognition. This can be achieved by having the system automatically surface only the most consequential issue in the decision-maker's argument, or by allowing decision-makers to conduct the quick-test themselves and bypassing reflection or skipping certain reflection steps when they judge the uncertainty to be low. System parameters 
can also be tuned to limit the number and complexity of issues presented. More broadly, varying these design choices enables \aact{} to be adapted to different users and contexts: high-stakes, time-insensitive, or analytically complex tasks, and users who can tolerate higher cognitive load, may benefit from richer, higher-demand configurations of \aact{}, whereas low-stakes, time-constrained, less complex tasks or novice decision-makers who may be more easily overwhelmed are better served by lighter-weight versions of \aact{} that emphasizes only the most essential reflections. 
Such adaptation can also operate at the level of
\textit{task instances}: for example, more difficult tasks---identified by AI's or human's low confidence---can employ the higher-demand version of \aact{}, while simpler tasks could use a lighter version.

\subsubsection{Additional design consideration for deploying \aact{}}
\label{sec:additionaldesign}
Beyond scalability-oriented adaptations, 
there exist additional factors that can further influence how \aact{} operates in real-world deployments. For example, in addition to the user-triggered quick-test described earlier, \textit{user control} can be expanded by allowing decision-makers to tune the system parameters that determine the granularity of counterfactual analysis, choose the level of details in AI-provided information (e.g., concise summaries vs. full explanations), set personal ``reflection budgets'' that cap the number of issues to reflect on for any task, or request custom counterfactual analysis.

Likewise, \textit{interface design} offers substantial opportunities to improve usability. Visualization aids such as bar charts or compact graphical summaries can replace or augment table-based triangulation to enable faster comparison without overloading users, and formative user studies may reveal more streamlined ways to present information. Designers may also explore more interactive layouts, such as modular panels or visual maps of argument quality, rather than relying on a fixed sequential dialogue. Although our study used pre-set AI message templates to ensure consistency and avoid hallucinations, future instantiations may incorporate LLMs to provide richer and more adaptive conversational scaffolding. For instance, LLMs could tailor the tone, styles and details of messages dynamically to match decision-makers' expertise while remaining grounded in domain-specific analysis. Such hybrid designs could preserve the reliability of template-based messaging while leveraging LLMs' flexibility to enhance engagement and accessibility.

\subsubsection{Designing AI Systems for Critical Thinking Beyond \aact{}}
While our work operationalizes \aact{} as an \textit{intervention} embedded directly within the decision making process, prior research on fostering critical thinking suggests several alternative design pathways that AI systems may explore. In particular, \aact{} resembles \textit{anchored instruction} in educational settings---where learners develop critical thinking through authentic, situated problems \cite{abrami2015, behar2011}---and parallels \textit{guided reflection} techniques used in clinical diagnosis \cite{Mamede2008, ilgen2013comparing, lambe2018guided, costa2019effects}, both of which view reflective scaffolding as most effective when integrated into problem-solving.
An alternative approach treats critical thinking development more \textit{pedagogically}, drawing on methods like \textit{mentoring} \cite{abrami2015} or \textit{programmatic training} \cite{behar2011}. Rather than intervening during the task, AI systems could employ \aact{}-like mechanisms as \textit{pre-task training} to build metacognitive skills before real decisions, which may be especially useful when opportunities for in-situ reflection are limited or when cognitive load must remain low during task execution.

Another consideration is whether critical thinking should be cultivated through \textit{domain-specific} or \textit{domain-general} instruction. The literature offers mixed evidence on the relative benefits of each \cite{Ennis1989, lai2011, abrami2008, bailin2002, behar2011}. Our current \aact{} system 
aligns with a \textit{domain-specific immersion} approach. Future work may explore AI systems that explicitly teach \textit{generalizable} critical thinking principles \cite{Ennis1989, bailin2002}, such as argument evaluation, evidence relevance, or bias detection, either alongside or independent from domain-specific reasoning.

\subsection{Ethical Considerations of \aact{}}

Designing AI systems that deliberately induce cognitive effort for critical reflection raises important ethical considerations. While Section~\ref{sec:design} shows that AACT's design choices can be adjusted to reduce burden, the ethical tension remains: deeper reflection often requires greater cognitive load, which risks cognitive overload \cite{cognitiveload}, elevates stress \cite{gjoreski2017monitoring, stress, alsuraykh2019stress, hou2015eeg}, and can impair performance \cite{biondi2021overloaded}. Transparency about these demands and giving users meaningful control over how deeply they engage are therefore essential, as is calibrating challenge to the level of ``desirable difficulties'' \cite{bjork2011making} that support metacognitive gains without imposing unnecessary strain. 

Another related concern is the self-doubt and cognitive dissonance that users may experience when being prompted to scrutinize their own reasoning. Although \aact{} is not designed to be antagonistic \cite{cai2024antagonistic} or argumentative \cite{musi2025toward}, repeated critiques 
can still evoke discomfort \cite{aronson1969theory}. Moderate dissonance can strengthen reflective thinking \cite{heyes2017enigma, deliu2025cognitive}, but excessive or poorly calibrated criticism risks undermining confidence or creating negative emotional experiences. 
Responsible deployment therefore requires sensitivity to users' affective states, cultural expectations around critique, and the context in which \aact{} is used.

Taken together, these ethical issues underscore the need for participatory evaluation with end-users to detect unintended emotional or cognitive consequences, and to guide refinements that keep \aact{} supportive rather than burdensome, ensuring that reflective scaffolding ultimately strengthens autonomy without compromising user well-being.

\subsection{Generalizability and Future Work}

We acknowledge that the generalizability of our results is limited by various choices of our study design. 
Therefore, evaluating the effectiveness and generalizability of \aact{} beyond the context of our case study is an important future work.

\noindent textbf{Task Choice: }
Our user study chooses house price prediction tasks as a case study, which limits the result generalizability.
In this task, participants often reported strong prior knowledge and high confidence in their own judgments; for example, to a survey question on whether the AI was helpful or not, one participant responded ``\textit{Not really helpful as I just bought a home 2 years ago. I know best.}'' This may help explain the low levels of AI reliance we observed across all treatments when compared to prior work.
Future research should evaluate \aact{} in tasks where humans have less prior knowledge to see if patterns of engagement and reliance may differ.

\noindent \textbf{Task Stakes: }
Our results are limited by the low-stake nature of house price prediction and the online crowdsourcing study context. 
Meanwhile, critical thinking research in high-stake domains (e.g.,
clinical decision-making) often involves participants with specialized expertise (e.g., medical students or residents) \cite{staal2022, Taro2013, o2019cognitive,kilian2019, Mamede2007, Mamede2008, Mamede2010, ilgen2013comparing, lambe2018guided, costa2019effects, mamede2023deliberate}. 
Thus, future work should evaluate \aact{} with different participants, higher-stake tasks, or deliberately increase stakes by imposing higher rewards, or even penalties.

\noindent \textbf{Data Modality: }
Another limitation lies in our choice of a task using tabular data.  
Because \aact{}'s counterfactual analysis perturbs features to examine changes in model confidence, the framework requires access to meaningful, interpretable input features.
While this holds naturally for tabular datasets, extending \aact{} to text, image, or multimodal datasets may require additional feature extraction.
Crucially, the extracted features should be human-understandable; relevant methods include interpretable embeddings, concept activation vectors, or human-understandable attributes~\cite{liu2023,koh2020concept,kim2018interpretability,ghorbani2019towards,achanta2012slic}.
We encourage future research to evaluate \aact{} beyond tabular data and explore how different feature extraction methods may influence how users construct arguments and interpret feedback.

\noindent \textbf{AI Model Performance: }
Our user study used a relatively well-performing AI model. 
However, high-performing AI models may be unavailable for certain tasks or infeasible for contexts where AI's cost and efficiency is prioritized.
Crucially, \aact{} relies primarily on AI model's probability function instead of decision performance, so future research can study how an AI model's performance impact its counterfactual analysis and AI-based correction suggestions.

\subsection{Additional Limitations}

Our study leaves several additional limitations. Although we evaluated \aact{} using standard human–AI decision making metrics, we did not directly assess participants' critical thinking skills. Future research could incorporate standardized critical-thinking assessments \cite{rear2019one} or embed reasoning problems  \cite{le2018} that explicitly measure metacognitive gains. Our study also examined \aact{} in a single-session setting; longitudinal studies are needed to understand how reasoning strategies evolve over time, and whether repeated exposure strengthens or attenuates the benefits observed. Moreover, while our implementation focused on individual decision-makers, extending \aact{} to group decision making could reveal new opportunities and challenges for reflective facilitation. Exploring these directions will help clarify the broader impact of \aact{} and refine its design for real-world use.

%% file: tables/benefit_costs.tex
\begin{table*}[th!]
\centering
\begin{tabular}{p{4cm} p{5.5cm} p{5.5cm}}
\toprule
\addlinespace[3pt]
\textbf{Dimension} & \textbf{Benefits} & \textbf{Costs} \\
\addlinespace[3pt]
\midrule
\addlinespace[3pt]
\textbf{Reliance behavior} 
& Reduced over-reliance when AI is wrong
& Increased under-reliance when AI is correct \\
\addlinespace[3pt]
\midrule
\addlinespace[3pt]
\textbf{Metacognitive engagement} 
& Stronger critique of completeness, reliability, and conflicts in reasoning 
& Higher cognitive load and mental effort \\
\addlinespace[3pt]
\midrule
\addlinespace[3pt]
\textbf{Decision quality} 
& Accuracy gains for the subgroup of decision-makers who self-reported to be very familiar with AI
& No accuracy improvement for the general population\\
\addlinespace[3pt]
\midrule
\addlinespace[3pt]
\textbf{Efficiency} 
& More deliberate, thoughtful decisions 
& More interaction steps and longer decision-making time\\
\addlinespace[3pt]
\midrule
\addlinespace[3pt]
\textbf{User experience} 
& Some improvement in perceived critical thinking abilities; preferred by decision-makers who value autonomy and reasoning transparency 
& Cognitively overloading and trigger self-doubt; received lowest subjective perceptions among decision-makers with lower education levels\\
\addlinespace[3pt]
\bottomrule
\end{tabular}
\vspace{2mm}
\caption{Benefits and costs of the current AACT system, based on user study results.}
\label{tab:aact_benefits_costs}
\vspace{-15pt}
\end{table*}

%% file: tables/design.tex
\begin{table*}[t!]
\small
\begin{center}
\begin{tabular}{p{3cm} p{6cm} p{6cm}}
\toprule
\addlinespace[4pt]
\textbf{Design Choices}& 
\textbf{Lower cognitive load} &
\textbf{Higher Cognitive Load} 
\\
\addlinespace[4pt]
\midrule
\addlinespace[4pt]
\textbf{Stages shown} &
Show only the stage corresponding to the most serious issue; additional stages revealed on demand
&
Present all four stages (agreement, incompleteness, unreliability, conflict) by default
\\
\midrule
\addlinespace[4pt]
\textbf{Steps within each stage}  &
Require only targeted self-reflection; other steps shown on demand
&
Full critique-and-correction workflow: 
\begin{enumerate}
    \item Targeted self-reflection
    \item AI-based correction suggestions
    \item Data-based triangulation
\end{enumerate}
\\
\addlinespace[4pt]
\midrule
\addlinespace[4pt]
\textbf{Bypass reflection and early stopping }&
User may bypass reflection entirely or stop early when they deem additional reflection as unnecessary&
User must complete all reflective steps with no option to bypass or stop early 
\\
\addlinespace[4pt]
\midrule
\addlinespace[4pt]
\textbf{System parameters} &
large $\epsilon$, small $k$, small \texttt{max\_feature\_change} to reduce the number and complexity of identified issues&
small $\epsilon$, large $k$, large \texttt{max\_feature\_change} to surface more issues with higher complexity
\\
\addlinespace[4pt]
\bottomrule
\end{tabular}
\end{center}
\caption{Different designs of \aact{} lead to different levels of cognitive load, and can be suitable for different types of users or contexts. }
\label{table:design}
\vspace{-15pt}
\end{table*}

%% file: sections/7_conclusion.tex
\section{Conclusions}
\label{sec:conclusion}
In this work, we introduced the AI-Assisted Critical Thinking (\aact{}) framework, a domain-grounded approach that supports human critical thinking in decision making by  
having AI perform counterfactual analysis on a decision-maker's decision argument,  surface potential flaws, and guide them through a structured critique-and-correction workflow. Our user study shows that \aact{} can reduce over-reliance on AI and strengthen decision autonomy, albeit with increased cognitive effort, revealing a key design trade-off for reflective AI systems. We further outline ways to improve \aact{}'s practical applicability, adapt it to different contexts, and address ethical considerations surrounding reflective scaffolding. Together, these contributions position \aact{} as an exploratory step towards building AI as partners in reasoning in human-AI decision making.

%% file: sections/appendix.tex
\vspace{10mm}
{\hspace{-3.5mm}\LARGE \textbf{Appendix}}

\input{sections/appendix/math}

\input{sections/appendix/ui}

\input{sections/appendix/study}

\input{sections/appendix/results}

\clearpage

\begin{figure*}[h!]
        \includegraphics[width=\textwidth]{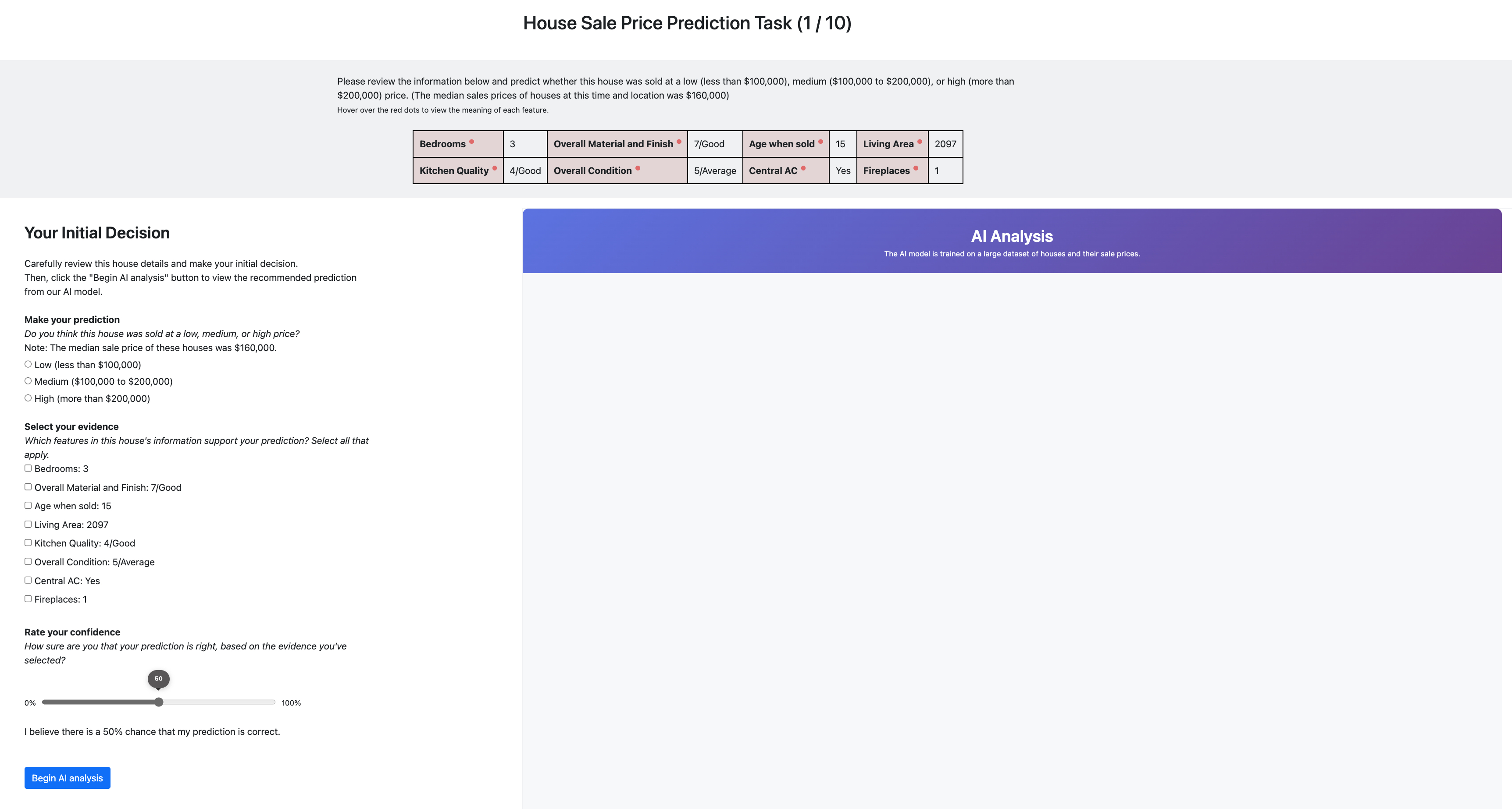}
    \caption{Example user interface for a decision-making task before AI assistance is shown, for the Recommender, Analyzer, and AACT treatments. AI assistance will be shown on the right after users submits their initial decision.}
    \label{fig:ui_aact0}
\end{figure*}

\begin{figure*}[h!]
    \centering
    \includegraphics[width=\textwidth]{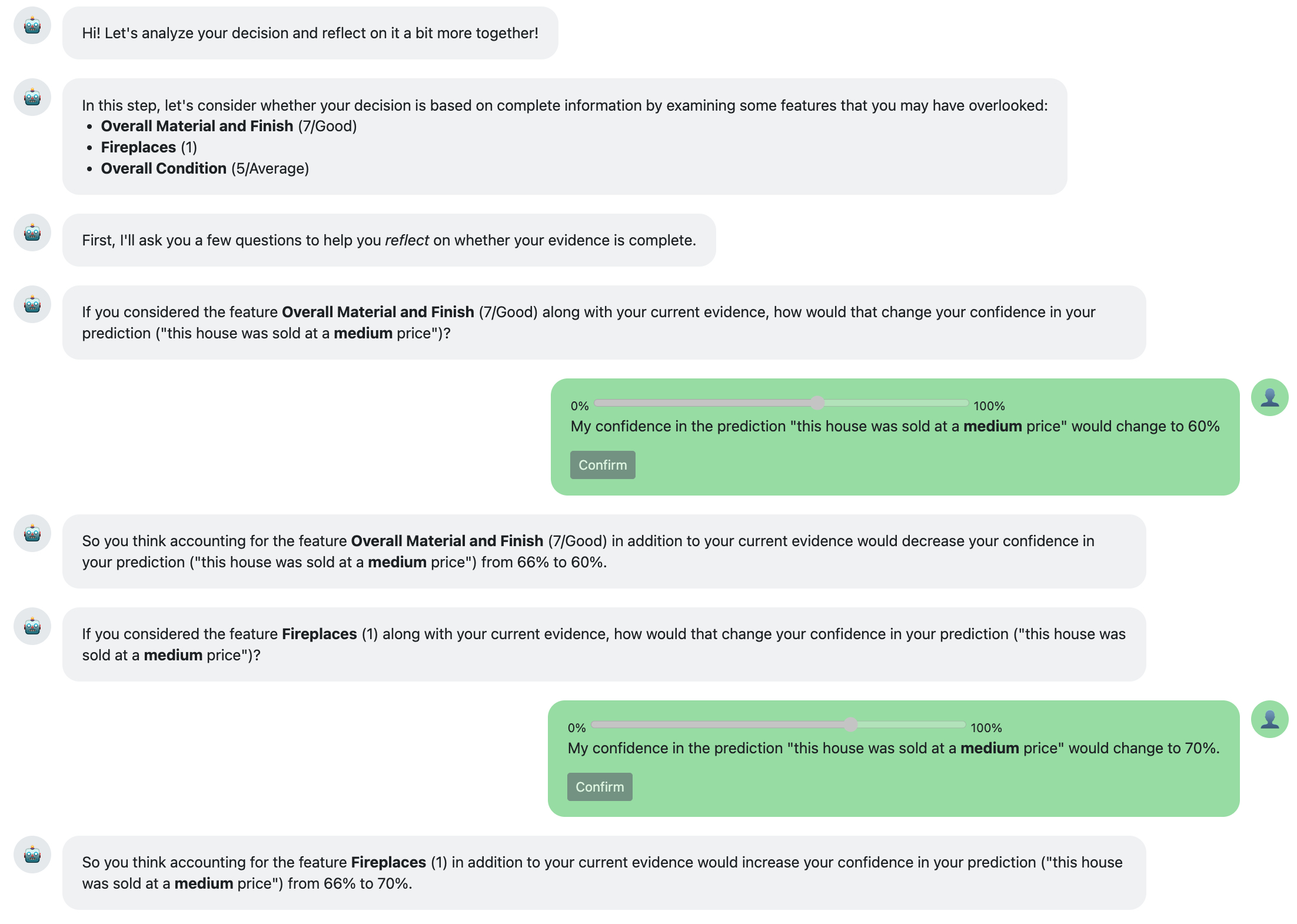}
    \caption{Example user interfaces for \aact{} during targeted self-reflection: user is asked questions about additional features and inputs their change in confidence using a slider.}
    \label{fig:ui_aact1}
\end{figure*}
\newpage
\begin{figure*}[h!]
    \centering
        \includegraphics[width=\textwidth]{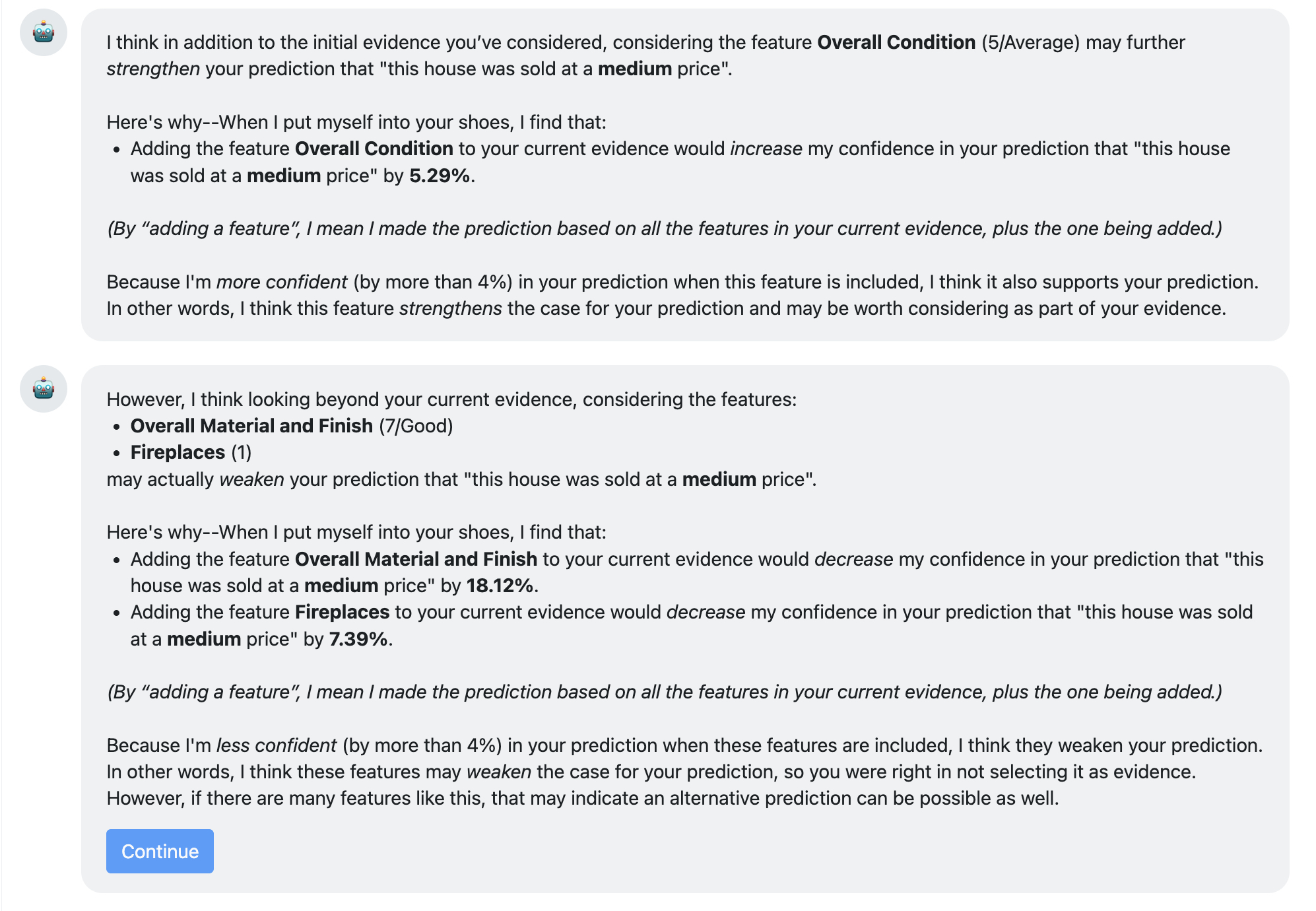}
        \caption{Example user interfaces for presenting AI-based correction suggestions in \aact{}: AI explains how adding certain features to the user's evidence might strengthen or weaken their prediction.}
        \label{fig:ui_aact2}
\end{figure*}
\newpage
\begin{figure*}[h!]
    \centering
        \includegraphics[width=\textwidth]{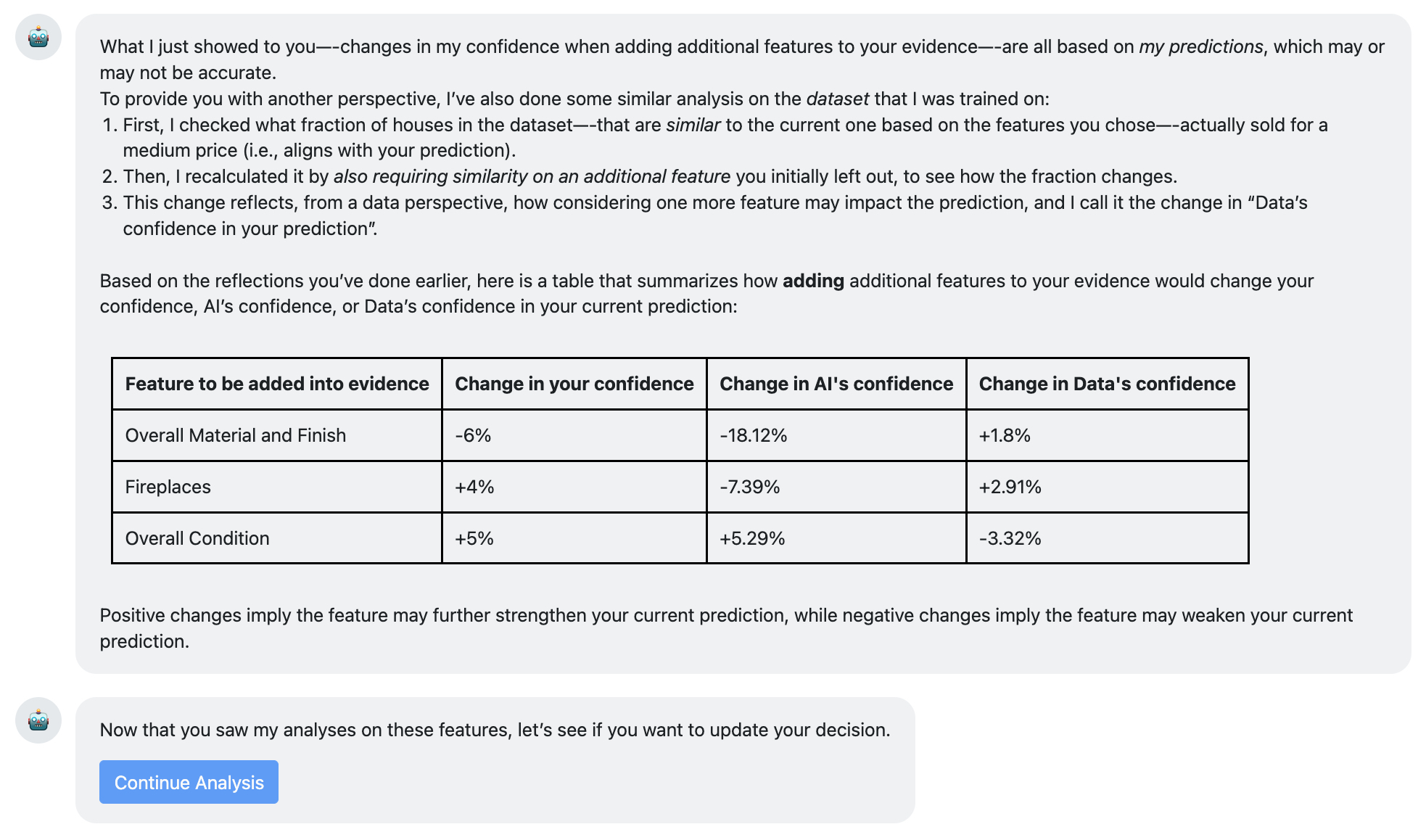}
        \caption{Example user interfaces for data-based triangulation in \aact{}: AI provides additional information from the empirical data distribution and explains how it was computed.}
        \label{fig:ui_aact3}
\end{figure*}
\newpage
\begin{figure*}[h!]
    \centering
        \includegraphics[width=0.9\textwidth]{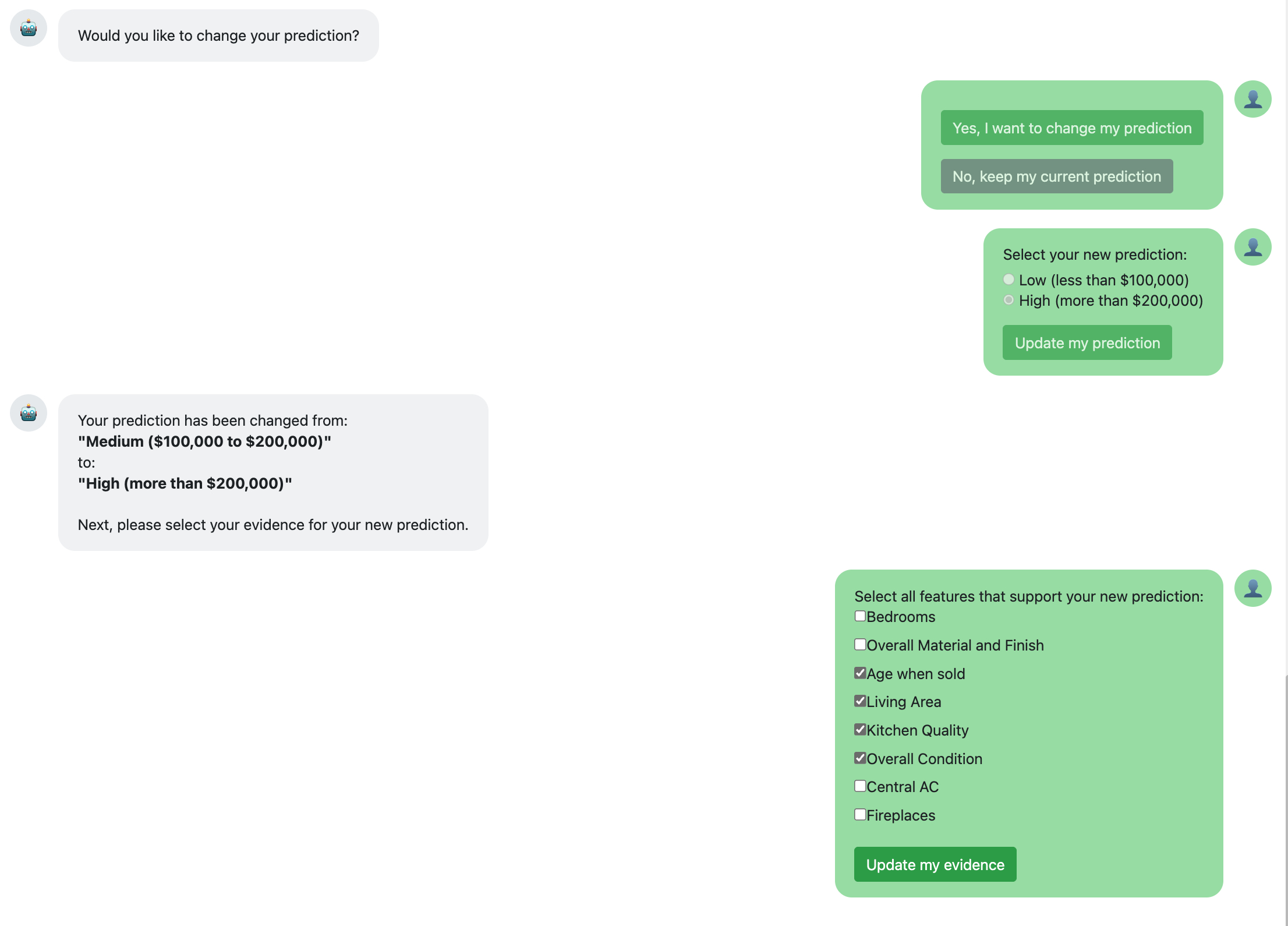}
        \caption{Example user interfaces for decision-makers to update their decision/argument: user is asked if they want to change their prediction, evidence, or confidence.}
        \label{fig:ui_aact4}
\end{figure*}


\input{tables/appendix/metrics}

\input{tables/appendix/demographics}

\input{tables/appendix/exit_survey}


\begin{figure*}[t]
    \centering
    \begin{subfigure}[b]{0.38\textwidth}
        \includegraphics[width=\textwidth]{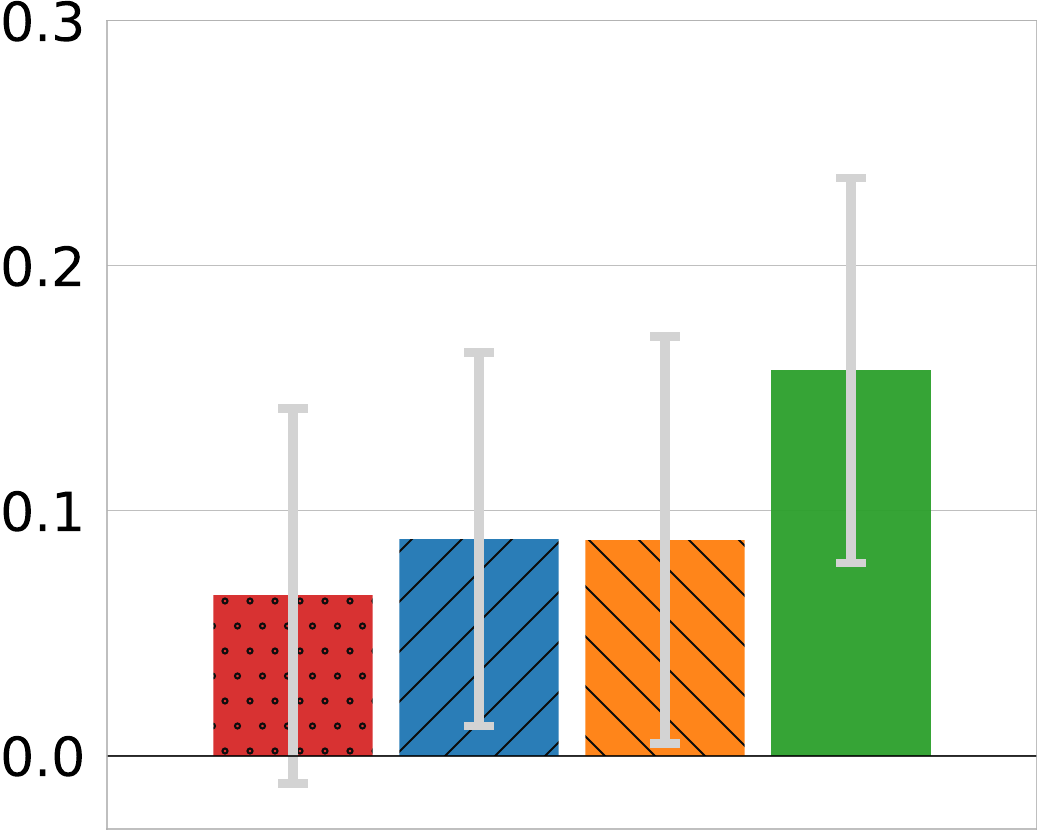}
        \caption{Learning during intervention}
        \label{fig:learningA}
    \end{subfigure}
    \hfill 
    \begin{subfigure}[b]{0.38\textwidth}
        \includegraphics[width=\textwidth]{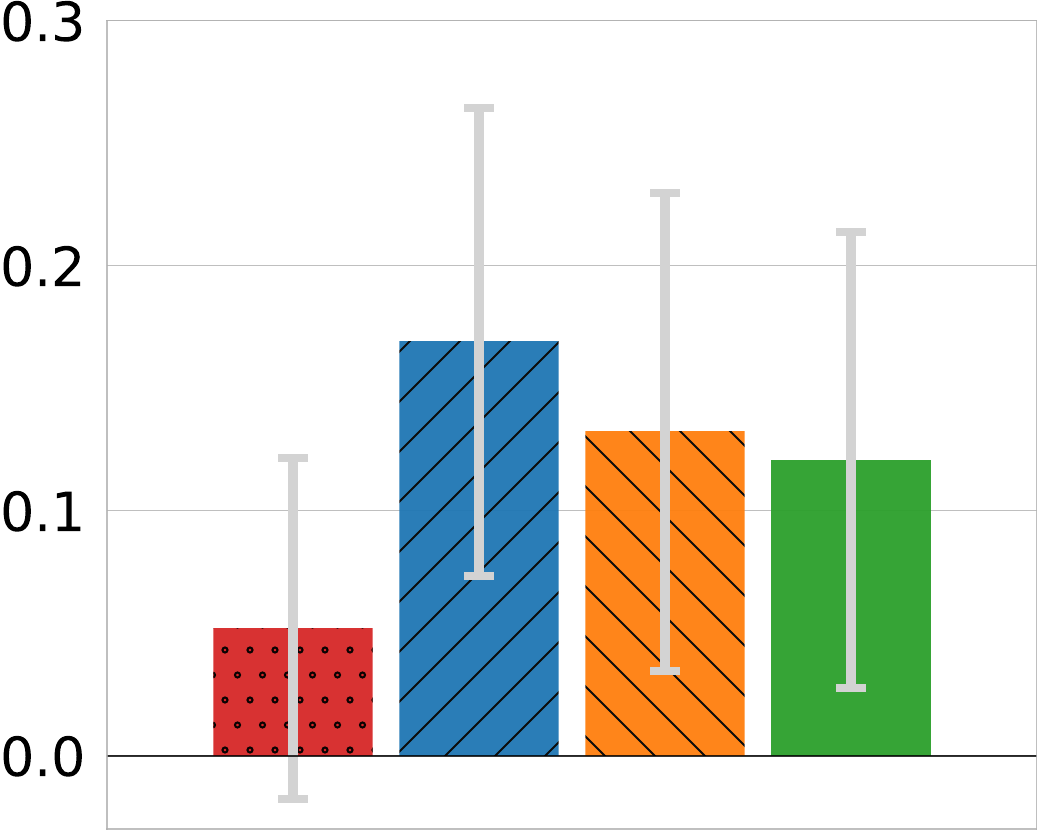}
        \caption{Learning after intervention}
        \label{fig:learningB}
    \end{subfigure}
    \begin{subfigure}[b]{0.2\textwidth}
        \includegraphics[width=\textwidth]{figs/figs_main/legends/legend_h.pdf}
        \vspace{12mm}
    \end{subfigure}
    \caption{Comparisons on participants' learning (a) while receiving the AI assistance intervention (b) after receiving the AI assistance intervention. 
    Values for the Human-only treatment are computed based on the normalized change of decision accuracy between corresponding sessions of tasks (learning during intervention: tasks 6--15 vs. tasks 1--5, learning after intervention: tasks 16--20 vs. tasks 1--5); they provide a baseline for organic learning happened due to repetitive task completion without AI assistance interventions. 
    Error bars represent 95\% confidence intervals of the mean values. }
    \label{fig:learning}
\end{figure*}

\begin{figure*}[h!]
    \centering
    \hspace{-5mm}
        \includegraphics[width=1.1\textwidth]{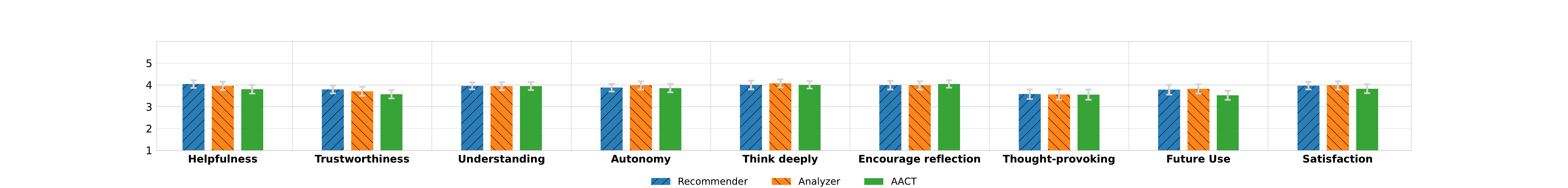}
    \caption{Comparisons on users' perceptions of AI across treatments. Error bars represent 95\% confidence intervals of the mean values.}
    \label{fig:survey2}
\end{figure*}

%% file: sections/appendix/math.tex
\section{Technical Details of \aact{}}
\label{app:math}

\subsection{Calculating AI's Confidence on Any Decision Given Any Argument}
\label{app:math1}
A critical building block for enabling AI's counterfactual perspective-taking is the estimation of $P_M(y^{*}|\boldsymbol{x^{*}})$---AI's confidence on any decision $y^{*}\in\mathcal{Y}$ given any argument $\boldsymbol{x^{*}} \subseteq \boldsymbol{x}$. 
Calculating $P_M(y^{*}|\boldsymbol{x^{*}})$ requires us to marginalize out the features not included in the argument $\boldsymbol{x^{*}}$, denoted as $\overline{\boldsymbol{x^{*}}}$ (i.e., $\boldsymbol{x^{*}} \cup \overline{\boldsymbol{x^{*}}} = \boldsymbol{x}$):

\begin{equation*}
P_M(y^{*}|\boldsymbol{x^{*}}) = \int P_M(y^{*}|\boldsymbol{x^{*}}, \overline{\boldsymbol{x^{*}}})\cdot P(\overline{\boldsymbol{x^{*}}}|\boldsymbol{x^{*}})d\overline{\boldsymbol{x^{*}}}
\end{equation*}

\noindent 

where $P_M(y^{*}|\boldsymbol{x^{*}}, \overline{\boldsymbol{x^{*}}})$ can be given by the AI model $m$ and $P(\overline{\boldsymbol{x^{*}}}|\boldsymbol{x^{*}})$ is the value distribution for features that are not a part of the argument $\boldsymbol{x^{*}}$ given the specified feature values in the argument. Since the exact integration is difficult to compute, we use Monte Carlo sampling to approximate this marginalized probability based on a set of $L=5000$ instances 
sampled from the empirical distribution of $P(\overline{\boldsymbol{x^{*}}}|\boldsymbol{x^{*}})$ given by the AI model's training dataset $D$:
\begin{equation*}
P_M(y^{*}|\boldsymbol{x^{*}}) \approx \frac{1}{L} \sum_{i=1}^{L} P_M(y^{*}|\boldsymbol{x^{*}}, \overline{\boldsymbol{x_i^{*}}}) \quad\quad \text{where} \;\; \overline{\boldsymbol{x_i^{*}}}\sim P_D(\overline{\boldsymbol{x^{*}}}|\boldsymbol{x^{*}})
\end{equation*}

\noindent 

Here, $P_D(\overline{\boldsymbol{x^{*}}}|\boldsymbol{x^{*}})$ is computed by restricting sampling from training data instances whose values on argument features are the same as those specified in $\boldsymbol{x^*}$. In practice, training data instances with exactly the same argument values as those specified in $\boldsymbol{x^*}$ may be sparse, potentially making the estimation of the empirical distribution  $P_D(\overline{\boldsymbol{x^{*}}}|\boldsymbol{x^{*}})$ unreliable. To address this issue, we may further simplify the approximation assuming the independence between argument features (i.e., features in $\boldsymbol{x^{*}}$) and non-argument features (i.e., features in $\overline{\boldsymbol{x^{*}}}$):
\begin{equation*}
P_M(y^{*}|\boldsymbol{x^{*}}) \approx \frac{1}{L} \sum_{i=1}^{L} P_M(y^{*}|\boldsymbol{x^{*}}, \overline{\boldsymbol{x_i^{*}}}) \quad\quad \text{where} \;\; \overline{\boldsymbol{x_i^{*}}}\sim P_D(\overline{\boldsymbol{x^{*}}})
\end{equation*}

\subsection{Generating Arguments for Alternative Decisions}
\label{app:math2}

To generate strong conflicting arguments, for each alternative decision $y_c\neq y_h$, AI can identify the ``strongest argument'' for that decision $\boldsymbol{x_c}= \arg\max_{\boldsymbol{x_c^{'}}\subseteq\boldsymbol{x}} P_M(y_c|\boldsymbol{x_c^{'}})$, which is a subset of features such that when focusing only on their values, 
AI's confidence in $y_c$ reaches the highest. In practice, searching through all possible feature subsets to identify $\boldsymbol{x_c}$ can be computationally challenging.
Thus, we use LIME, a popular post-hoc explainable AI method, to explain the AI model $m$ and approximate the computation of $\boldsymbol{x_c}$. 
LIME can generate feature importance score $s(X_i, y^*)$ for every feature $X_i$ towards any decision $y^*$.
When constructing $\boldsymbol{x_c}$ for a specific alternative decision $y_c$, we can rank all features based on their ``contribution'' towards $y_c$ using their feature importance score $s(X_i, y_c)$.
Then, we include only the ones with $s(X_i, y_c)>\mu$ ($\mu>0$ is a pre-specified threshold) into $\boldsymbol{x_c}$.

%% file: sections/appendix/ui.tex
\section{Example AACT User Interfaces}
\label{app:UI}

Figure~\ref{fig:ui_aact0} shows an example user interface for a decision-making task before AI assistance is shown, for the Recommender, Analyzer, and AACT treatments. AI assistance will be shown on the right after users submits their initial decision.
Figures~\ref{fig:ui_aact1}, ~\ref{fig:ui_aact2}, ~\ref{fig:ui_aact3},  ~\ref{fig:ui_aact4},  show example user interfaces for the \aact{} treatment.

%% file: sections/appendix/study.tex
\section{Additional User Study Details}
\label{app:study}

\subsection{Dataset and AI Model Details}
\label{app:dataset}
\textbf{Dataset: }
We used the Ames Housing Dataset, which contained 79 features and the regression problem was to predict house sales price. We selected 8 features: \textit{number of bedrooms, number of central AC, number of fireplaces, overall material and finish, kitchen quality, overall condition, age when sold, living area}. All features except the \textit{age when sold} feature were used as they were; the \textit{age when sold} feature was constructed by subtracting each house's \textit{Year Sold} feature by \textit{Year Built} feature.
To generate discrete class labels, we converted the house sales price into three categories, {\em Low} (less than \$100,000), {\em Medium} (\$100,000 to \$200,000), and {\em High} (more than \$200,000), resulting in an imbalanced dataset with 237, 1837, and 857 instances in each class and 2930 instances in total.

\noindent \textbf{AI Model: }
We divided the dataset into the training and test set based on a 80:20 split. We performed a grid search on the Logistic Regression model using the training set only and used the best-performing set of hyperparamaters: we used the Newton-CG (conjugate gradient) solver while all other hyperparameters were set to Scikit-learn's default values.
The model's performance on the test set was: accuracy=0.8737, balanced accuracy$=0.795$, F1 score for class \textit{Low}=$0.694$, F1 score for class \textit{Medium}=$0.902$, F1 score for class \textit{High}$=0.851$.

\subsection{Metrics and Questions}
Table~\ref{table:metric} shows the metric details for reliance and learning measurements. 

Table~\ref{table:demo} shows the full list of questions in the demographic survey. 
Education data was collected using the question ``What is your highest level of education completed?'' and the options:\textit{ Prefer not to answer, No formal education, Elementary school, Middle school, High school diploma or equivalent (GED), Some college no degree, Trade school or certificate program, Associate's degree, Bachelor's degree, Master's degree, Professional degree, Doctorate or higher}.
                            
Table~\ref{table:survey} shows the full list of questions in the exit survey.

\subsection{Demographics data}
\label{app:demo}

\textbf{Data from demographics survey: } 
Task familiarity was measured using the question ``How familiar are you with the task background of real estate markets?'' and a 5 point Likert scale. The participants' distribution was: $1:n=90,2:n=105,3:n=112,4:n=77,5:n=18$.
AI familiarity was measured using the question ``How familiar are you with artificial intelligence (AI) technologies?'' and a 5 point Likert scale. The participants' distribution was: $1:n=2,2:n=31,3:n=83,4:n=193,5:n=93$.
The participants' education distribution was: 
Prefer not to answer: $n=1$, 
No formal education: $n=0$, 
Elementary school: $n=0$, 
Middle school: $n=3$, 
High school diploma or equivalent (GED): $n=39$, 
Some college no degree: $n=74$, 
Trade school or certificate program: $n=13$, 
Associate's degree: $n=32$, 
Bachelor's degree: $n=147$, 
Master's degree: $n=76$, 
Professional degree: $n=5$, 
Doctorate or higher: $n=12$.

\noindent \textbf{Data from Prolific:}
We also report demographics distribution provided by Prolific, including their age, gender, and ethnicity:
\begin{itemize}
    \item [Age:] $M=42, SD=13.02$
    \item [Gender:] female: $n=210$, male: $n=200$
    \item [Ethnicity:] White: $n=295$, Black: $n=58$, Asian: $n=28$, Mixed: $n=18$, Other: $n=12$, Data expired: $n=2$
\end{itemize}

%% file: sections/appendix/results.tex
\section{Additional User Study Results}
\label{app:results}

\subsection{RQ2: Effects on Learning}
\label{app:RQ2}

To understand if \aact{} helps decision-makers build knowledge about the decision domain that are necessary for them to solve future decision tasks on their own, we look into participants' learning while they were actively receiving AI assistance (i.e., the improvement of their initial decision accuracy in the intervention stage over their decision accuracy in the pre-test) and their learning upon the completion of the AI assistance (i.e., the improvement of decision accuracy in post-test over that in pre-test; also referred to as the ``incidental learning''~\cite{Gajos2022}). As shown in Figure~\ref{fig:learning}, there is no significant difference across treatments for both the learning during intervention ($F(3,398)=1.046, p=0.372$) and learning after intervention ($F(3,398)=1.193, p=0.312$).  

Nonetheless, we note an interesting trend that, compared to other treatments with AI assistance, \aact{} seemed to result in more learning when participants were actively receiving the AI assistance ($M=0.157,SD=0.404$), but such learning did not seem to fully transfer to the post-test stage. 
We conjecture that one possible reason might be that the \aact{} treatment was generally more time-consuming and mentally demanding than the other two treatments with AI assistance (as shown in Section~\ref{sec:rq3}), thus participants in the \aact{} treatment may be more fatigued in the post-test stage, limiting their ability to show optimal independent decision performance.

\subsection{RQ3: Effects on Perceptions of AI}
\label{app:RQ3}
Figure ~\ref{fig:survey2} shows participants' survey metrics on perceptions of AI assistance; One-way ANOVA tests yielded no significant differences between treatments.

\subsection{RQ4: Individuals with lower education has more negative subjective perceptions of AACT}
\label{app:RQ4}

In general, we find participants with lower education levels (e.g., without bachelors) have more negative perceptions of \aact{}.  For example, within the subgroup of participants with no bachelor degrees, we detect significant differences in their perceptions of 
AI's trustworthiness ($F(2,114)=3.959, p=0.022, \eta^2=0.065$) and future use of AI ($F(2,114)=5.311, p=0.006, \eta^2=0.085$) across treatments---those who used \aact{} indicated lower trustworthiness of the AI assistance 
($M=3.184, SD=1.01$) 
than those used Recommender ($M=3.725, SD=0.905$; $p=0.034
, d=-0.565$), and \aact{} users also reported lower willingness for future use of the AI ($M=3.184, SD=1.036$) compared to 
Recommender users ($M=3.725, SD=0.905$; $p=0.034, d=-0.565$) and Analyzer users ($M=3.872, SD=0.978$; $p=0.012, d=-0.683$). 
Furthermore, 
when analyzing within participants of the \aact{} treatment, we also found those without bachelors reported significantly lower scores on all AI perception metrics compared to those with bachelors ($p<0.001$).

%% file: tables/appendix/metrics.tex
\begin{table*}[t!]
\centering
\small

\begin{tabular}{p{0.07\linewidth}p{0.2\linewidth}p{0.56\linewidth}}
\toprule
\textbf{Category} & \textbf{Measurement} & \textbf{Details} \\
\midrule

\multirow{6}{*}{Reliance} &
Agreement Fraction &
$\frac{\text{Number of tasks where humans' final decisions match AI's prediction}}{\text{Total number of tasks}}$ \\[1mm]
\cmidrule(lr){2-3}
& 
Switch Fraction &
$\frac{\text{Number of tasks where humans switch to match AI in their final decisions}}{\text{Total number of tasks where humans' initial decisions differ from AI}}$ \\[1mm]
\cmidrule(lr){2-3}
& 
Over-reliance Ratio &
$\frac{\text{Number of tasks where AI is wrong and humans' final decisions match AI}}{\text{Total number of tasks where AI is wrong}}$\\[1mm]
\cmidrule(lr){2-3}
& 
Under-reliance Ratio &
$\frac{\text{Number of tasks where AI is correct and humans' final decisions differ from AI}}{\text{Total number of tasks where AI is correct}}$ \\[1mm]
\midrule

\multirow{11}{*}{Learning} &
\multirow{6}{*}{Learning during intervention} & 
\begin{equation*}
    \text{Normalized change} = 
    \begin{cases}
        \frac{\textit{acc(intervention)} - \textit{acc(pre-test)}}{1 - \textit{acc(pre-test)}} & \textit{if acc(intervention)} > \textit{acc(pre-test)} \\
        \frac{\textit{acc(intervention)} - \textit{acc(pre-test)}}{\textit{acc(pre-test)}} & \textit{if acc(intervention)} < \textit{acc(pre-test)} \\
        0 & \textit{otherwise}
    \end{cases}
\end{equation*}
\\
\cmidrule(lr){2-3}& 
\multirow{6}{*}{Learning after intervention} & 
\begin{equation*}
    \text{Normalized change} = 
    \begin{cases}
        \frac{\textit{acc(intervention)} - \textit{acc(pre-test)}}{1 - \textit{acc(pre-test)}} & \textit{if acc(intervention)} > \textit{acc(pre-test)} \\
        \frac{\textit{acc(intervention)} - \textit{acc(pre-test)}}{\textit{acc(pre-test)}} & \textit{if acc(intervention)} < \textit{acc(pre-test)} \\
        0 & \textit{otherwise}
    \end{cases}
\end{equation*}
\\
\bottomrule
\end{tabular}
\caption{Metric details for reliance and learning measurements.}
\label{table:metric}
\end{table*}

%% file: tables/appendix/demographics.tex
\begin{table*}[h!]
\centering
\small

\begin{tabular}{p{0.2\linewidth}p{0.74\linewidth}}
\toprule
\textbf{Category} & \textbf{Questions} \\
\midrule

\multirow{4}{*}{Need for Cognition (NFC)} &
I would prefer complex to simple problems.\\
\cmidrule(l){2-2}
& I like to have the responsibility of handling a situation that requires a lot of thinking.\\
\cmidrule(l){2-2}
& \textbf{*}Thinking is \textbf{not} my idea of fun.\\
\cmidrule(l){2-2}
& \textbf{*}I would rather do something that requires little thought than something that is sure to challenge my thinking abilities.\\
\midrule

AI Familiarity &
How familiar are you with artificial intelligence (AI) technologies?\\
\midrule

Task Familiarity &
How familiar are you with the task background of real estate markets?\\
\midrule

\multirow{3}{*}{Cognitive Reflection Test (CRT)} &
A bat and a ball cost \$1.10 in total. The bat costs \$1.00 more than the ball. How much does the ball cost?\\
\cmidrule(l){2-2}
& If it takes 5 machines 5 minutes to make 5 widgets, how long would it take 100 machines to make 100 widgets?\\
\cmidrule(l){2-2}
& In a lake, there is a patch of lily pads. Every day, the patch doubles in size. If it takes 48 days for the patch to cover the entire lake, how long would it take for the patch to cover half of the lake?\\
\bottomrule
\end{tabular}

\caption{Demographics survey questions, excluding education.\textbf{*} indicated questions with reverse scoring.}
\label{table:demo}
\end{table*}

%% file: tables/appendix/exit_survey.tex
\begin{table*}
\centering
\small

\begin{tabular}{p{0.16\linewidth}p{0.16\linewidth}p{0.62\linewidth}}
\toprule
\textbf{Category} & \textbf{Measurement} & \textbf{Questions} \\
\midrule

\multirow{4}{*}{User Experience} &
Decision Confidence &
I feel confident in the decisions I made. \\
\cmidrule(lr){2-3}
& Efficacy &
I am confident about my ability to complete the decision making task. \\
\cmidrule(lr){2-3}
& Mental Demand &
The process of making predictions is mentally demanding. \\
\cmidrule(lr){2-3}
& Effort &
I have to work hard (mentally and physically) to accomplish my level of performance. \\
\midrule

\multirow{4}{*}{Critical Thinking} &
Evaluate Evidence &
In each task, before reaching my final decision, I evaluated all evidence thoroughly. \\
\cmidrule(lr){2-3}
& Multiple Perspectives &
I considered each prediction from multiple perspectives and examined multiple hypotheses. \\
\cmidrule(lr){2-3}
& Counter Evidence &
When working on each prediction task, I actively seek evidence that might counter what I already know. \\
\cmidrule(lr){2-3}
& Explain Prediction &
For each prediction, I can give reasons and arguments for my decision. \\
\midrule

\multirow{9}{*}{Perceptions of AI} &
Helpfulness &
The AI model's assistance is helpful for me to make good decisions. \\
\cmidrule(lr){2-3}
& Trustworthiness &
The AI model can be trusted to provide reliable decision support. \\
\cmidrule(lr){2-3}
& Understanding &
I understand how the AI model works to help me make predictions. \\
\cmidrule(lr){2-3}
& Autonomy &
My interaction with the AI model in these tasks helped me maintain the control to make predictions in the way I see fit. \\
\cmidrule(lr){2-3}
& Think Deeply &
The AI model helped me think more deeply about my decisions. \\
\cmidrule(lr){2-3}
& Encourage Reflection &
The AI model encouraged me to reflect critically on my initial assumptions. \\
\cmidrule(lr){2-3}
& Thought-provoking &
I found the interaction with the AI model thought-provoking. \\
\cmidrule(lr){2-3}
& Future Use &
I would like to continue working with the AI model. \\
\cmidrule(lr){2-3}
& Satisfaction &
I am satisfied with the AI model's assistance and the decision-making process. \\
\midrule

\multirow{3}{*}{Open-ended Questions} &
---- & For the last 5 tasks you completed, when AI assistance was not available, describe your general decision strategies. \\
\cmidrule(lr){2-3}& 
---- & Why did you find the AI assistance helpful or not helpful for making decisions? How can it be improved? \\
\cmidrule(lr){2-3}& 
---- & Which AI assistance do you prefer and why? (1) This study’s assistance without showing the AI’s prediction, or (2) assistance that directly shows the AI’s prediction. \\
\bottomrule
\end{tabular}

\caption{Exit survey questions. User Experience, Critical Thinking, and Perceptions of AI questions are measured using Likert scales.}
\label{table:survey}

\end{table*}